\documentclass[twocolumn]{aastex631}
\usepackage{amsmath,amssymb,bm}

\def \be{\begin{equation}}
\def \ee{\end{equation}}
\def \bea{\begin{eqnarray}}
\def \eea{\end{eqnarray}}
\def \la{\langle}
\def \ra{\rangle}

\hypersetup{
    urlcolor=cyan,
    }

\shorttitle{Flux transport in a RIAF}
\shortauthors{Dhang, Bai \& White}

\graphicspath{{./}{figures/}}
\begin{document}

%\title{Magnetic flux transport in a RIAF and the MAD connection}

\title{Magnetic Flux Transport in Radiatively Inefficient Accretion Flows and the Pathway towards a Magnetically Arrested Disk}

\email{prasundhang@gmail.com}
\email{xbai@tsinghua.edu.cn}
\author[0000-0001-9446-4663]{Prasun Dhang}
\affiliation{Institute for Advanced Study, Tsinghua University, Beijing 100084, China}
\affiliation{IUCAA, Post Bag 4, Ganeshkhind, Pune, Maharashtra 411007, India}
\affiliation{JILA, University of Colorado and National Institute of Standards and Technology, 440 UCB, Boulder, CO 80309-0440, USA}

\author[0000-0001-6906-9549]{Xue-Ning Bai}
\affiliation{Institute for Advanced Study, Tsinghua University, Beijing 100084, China}
\affiliation{Department of Astronomy, Tsinghua University, Beijing 100084, China}

\author{Christopher J. White}
\affiliation{Department of Astrophysical Sciences, Princeton University, Princeton, NJ 08544}

\begin{abstract}
Large-scale magnetic fields play a vital role in determining the angular momentum transport and generating jets/outflows in the accreting systems, yet their origin remains poorly understood. We focus on radiatively inefficient accretion flows (RIAFs) around the black holes (BHs), and conduct three-dimensional general-relativistic magnetohydrodynamic (GRMHD) simulations using the Athena++ code. We first re-confirm that the MRI dynamo in the RIAF alone
does not spontaneously form a magnetically arrested disk (MAD), conducive for the strong jet formation. We next investigate the other possibility, where the large-scale magnetic fields are advected inward from external sources (e.g. the companion star in X-ray binaries, magnetized ambient medium in AGNs).
Although the actual configuration of the external fields could be complex and uncertain, they are likely to be closed.
As a first study, we treat them as closed field loops of different sizes, shapes and field strengths.
Unlike earlier studies of flux transport, where magnetic flux is injected in the initial laminar flow, we injected the magnetic field loops
in the quasi-stationary turbulent RIAF in inflow equilibrium and followed their evolution. We found that a substantial fraction ($\sim15\%-40\%$) of the flux injected at the large radii reaches the BH with a weak dependence on the loop parameters except when the loops are injected at high latitudes, away from the mid-plane.
Relatively high efficiency of flux transport observed in our study hints that a MAD might be formed relatively easily close to the BH, provided that a source of the large-scale field exists at the larger radii.

\end{abstract}


\section{Introduction}
\label{sect:intro}
Astrophysical accretion disks influence the systems over a large range of scales spanning from  planet formation to galaxy evolution. 
They also energise the most powerful sources in the Universe. For example, disks orbiting around the stellar-mass black holes (BHs) and neutron stars are considered to be among the most luminous X-ray sources in the sky (\citealt{Remillard2006}). Active galactic nuclei (AGNs), powered by the accretion of matter onto a supermassive black hole at the centre of galaxies, are not only the most powerful sources but also energy released by the AGNs provides feedback to the entire galaxy and determines its evolution (\citealt{Silk_Rees_1998,Harrison2017,Morganti2017}). 

Broadly speaking, accretion occurs via three different modes: (i) geometrically thin and optically thick Keplerian disk (standard disk; \citealt{Shakura1973, Novikov1973}), (ii) geometrically thick and optically thin radiatively inefficient accretion flows (RIAF; \citealt{Chakrabarti1989, Narayan1994, Blandford1999}) and (iii) geometrically and optically thick slim disks (\citealt{Abramowicz1988}).
Slim disks accrete matter at a super-Eddington  rate, while mass accretion rate $\dot{m}$ is sub-Eddington in both the standard disk ($10^{-4} \lesssim \dot{m} / \dot{M}_{\rm Edd} \lesssim 1$) and in RIAFs ($\dot{m} / \dot{M}_{\rm Edd} \lesssim 10^{-4}$). In this paper, we focus on the RIAFs, where disks likely span most of their time (e.g., \citealt{Yuan2014}, such as disks around Sgr A$^*$ and the SMBH in M87), and their dynamics are relatively simple compared to the other two states.

The structure and evolution of the rotationally supported accretion disks 
are primarily determined by the process of angular momentum transport. The current consensus is that the magnetorotational instability (MRI; \citealt{Balbus1991}) gives rise to 
angular momentum transport and vigorous turbulence in a fully ionised accretion flow (e.g. in X-ray binaries, inner part of the AGN disk, sufficiently ionised part of proto-planetary disks). The MRI becomes more efficient in angular momentum transport if the accretion disk is threaded by a net vertical magnetic flux. This has been observed in the local shearing-box simulations that the presence of a net vertical magnetic flux enhances the MRI turbulence and hence angular momentum transport (\citealt{Bai2013}). Additionally, net flux threading the disk helps to launch the  winds/outflows (\citealt{Bai2013, Suzuki2014}).

A large-scale magnetic field close to the central accretor (a BH or a neutron star) is a necessary ingredient for jet production in accreting systems (\citealt{Blandford1977, Blandford1982}). It has been proposed that a RIAF saturated with strong poloidal magnetic flux close to the BH provides an ideal condition for jet production (\citealt{Bisnovatyi-Kogan1974a,Esin1997, Fender1999a, Narayan2003, Meier2005c}). The idea has been  verified in numerical simulations (\citealt{Igumenshchev2003, Narayan2012}). The studies found that a strongly magnetized RIAF, namely a magnetically arrested disk (MAD) around spinning BH produces strong jets, extracting net energy from the BH spin via the Penrose-Blandford-Znajek process (\citealt{Tchekhovskoy2011, McKinney2012}).

The MAD model predicts a correlation among the mass accretion rate, magnetic flux threading the BH and jet power which is found to be in agreement with the observations of radio-loud AGNs (\citealt{Zamaninasab2014,Ghisellini2014}). Recent polarization studies of M87 at 230 GHz from Even Horizon Telescope EHT observations (\citealt{EHT2021a, EHT2021b, Yuan2022}) also infer the presence of a dynamically important near-horizon organized, poloidal magnetic flux consistent with GRMHD models of MAD.

What could be the possible source of the magnetic flux close to the BH?
Most of the numerical simulations of MAD start with a strong enough large-scale poloidal flux which is eventually brought close to the BH and gets accumulated by flux-freezing (\citealt{Tchekhovskoy2011, McKinney2012}). However, the source of the large-scale field is not entirely obvious (see also \citealt{Begelman2022}).  It can potentially be generated in the disk itself by a dynamo action (\citealt{Bugli2014,Vourellis2021,Mattia2020, Mattia2022}) or be advected in from some external sources (\citealt{Cao2011, Li2021}). 
 
The efficiency of the dynamo action in generating coherent and strong large-scale poloidal field required to produce strong jets is found to be different for different numerical simulations of RIAFs. The $\alpha-$ effect (responsible for the generation of the poloidal field by a dynamo) is found to be weak  in simulations that start with small poloidal magnetic loops (\citealt{Hogg2018, Dhang2019, Dhang2020}). The quasi-stationary  states of those simulations  are in a weakly magnetized regime, popularly known as ``standard and normal evolution" (SANE; \citealt{Narayan2012}). However, recent simulations with a very strong (with gas to magnetic pressure ratio $\beta \approx 5$) and coherent initial toroidal field showed the production of large-scale poloidal field loops of the size of scale-height $H\propto R$ and led to the MAD eventually (\citealt{Liska2020}). Therefore, it is worth noting that the simulations need to start with a strong and  coherent large-scale field (either poloidal or toroidal) to achieve MAD.

In addition to the in-situ generations of the magnetic field by a dynamo process, it might be possible that an initially weak field supplied to the disk (from the outer part of the disk or companion star in case of XRBs, ambient medium of the AGNs) can in principle be amplified by flux freezing. Flux accumulation near the BH depends on the relative efficiency between the inward advection by the accretion flow and the outward diffusion due to a turbulent resistivity (\citealt{Lubow1994}). Additionally, turbulent pumping can also cause outward transport of the large-scale magnetic field in a dynamo-active accretion flow (\citealt{Dhang2020}).
However, a few studies proposed that vertical magnetic field accretion can be efficient in the hot, tenuous surface layer (coronal region, where radial velocity is comparatively higher compared to that in the mid-plane) in a hot accretion flow (\citealt{Beckwith2009}). 
It is also interesting to note that the simulations of large-scale accretion flow around the galactic centre fed by the magnetized winds of Wolf-Rayet stars also show efficient inward transport of magnetic field towards the centre (\citealt{Ressler2020a,Ressler2020b}).

This paper studies the magnetic flux transport in a fully turbulent RIAF, unlike the previous studies where magnetic flux is injected in the initial laminar condition. Therefore, first, we run a simulation to attain a quasi-stationary RIAF in the SANE regime (weakly magnetized). Then we inject the external magnetic flux on top of the existing magnetic field in this turbulent SANE RIAF. In the later part of the paper, we refer to it as the Initial RIAF run. It is customary to use net vertical magnetic flux threading the disk to investigate the flux transport (\citealt{Beckwith2009,Zhu2018, Mishra2019a}. However, we argue that the geometry of the external magnetic field is likely to be closed. In this paper, as a first step, we use magnetic field loops as the simplest possible form of the external magnetic field, studying its transport in the turbulent RIAF and its possibility of saturating the BH with magnetic flux towards the MAD regime.

We use general-relativistic MHD (GRMHD) simulations to study the magnetic flux transport. The usage of the general-relativistic  approach is crucial in our work. Many of the important diagnostics we used in our work involve computing fluxes (mass, angular momentum, magnetic flux) at the event horizon of the BH.  However, Newtonian MHD suffers from the effects of inner boundary conditions, which artificially affect the evolution of flow and  magnetic flux close to the BH. This is  avoided in GRMHD by placing the inner boundary within the event horizon so that the computation domain is causally disconnected from the inner boundary. We also neglect radiation physics in our GRMHD simulations as radiation is supposed to  play an insignificant role in determining the dynamics of RIAFs of low accretion rate (also, see \citealt{Dexter2021}).

The paper is organized as follows. In Section 2, we discuss the solution method and physical set-up of the  RIAF simulations. In Section 3, we discuss the evolution of the flow, convergence and magnetic state of the Initial RIAF run. We describe the method of flux injection and its results in section 4. Finally, the key points of results are discussed and summarized in Sections 5 and 6.

\section{Method}
\label{sect:method}
We performed two sets of simulations. In the first set, we performed a simulation to achieve a fully turbulent  quasi-stationary RIAF that forgets the initial condition. This is labelled as the Initial RIAF run. In the second set of simulations, we restart the Initial RIAF run at late times, inject external magnetic flux and  study its evolution for different parameters associated with injected magnetic field loops. In this section, we describe the formulation and simulation setup for all the simulations. Here we note that the initial condition of the Initial RIAF run is specified in this section. The parameters related to the restart setup will be described in section \ref{sect:results_flux}.  

\subsection{Equations solved}
\label{sect:method_eqs}
We solve the ideal general relativistic magnetohydrodyamic (GRMHD) equations
\bea
\label{eq:mass}
 && \partial_0 \left(\sqrt{-g} \rho u^0 \right) + \partial_j \left (\sqrt{-g} \rho u^j \right) = 0\\
  \label{eq:mass_energy}
 &&    \partial_0 \left(\sqrt{-g} T^{0}_{\mu} \right) + \partial_j \left( \sqrt{-g} T^{j}_{\mu} \right) = \frac{1}{2}\sqrt{-g} T^{\nu \sigma} \partial_{\mu}  g_{\nu \sigma} \\
     \label{eq:maxwell}
   && \partial_0 \left(\sqrt{-g} B^{i} \right) + \partial_j \left( \sqrt{-g} \  F^{*ij} \right) = 0 \\
   \label{eq:monopole}
   && \frac{1}{\sqrt{-g}} \partial_{i} \left( \sqrt{-g} B^{i} \right) = 0
\eea
in a spherical-like Kerr-Schild coordinates ($t,\ r, \ \theta, \ \phi$) with $G=c=M_{BH}=1$. All the length scales and time scales  in this work are expressed in units of the gravitational radius $r_g=GM_{BH}/c^2$ and $t_g=r_g/c$ respectively unless stated otherwise. Here, $g_{\mu \nu} $ and $g$  are metric coefficients and metric determinant  respectively. Following convention, Greek indices run through [0,1,2,3], while $i$ denotes a spatial index.  Equations \ref{eq:mass}, \ref{eq:mass_energy}, \ref{eq:maxwell} \& \ref{eq:monopole} describe conservation of particle number, conservation of energy-momentum, source-free Maxwell equations and no-magnetic monopole constraint respectively. Here 
 \be
 \label{eq:stress_tensor}
     T^{\mu \nu } = \left(\rho h + b^2 \right) u^{\mu} u^{\nu} + \left( p_{\rm gas} + \frac{b^2}{2} \right) g^{\mu \nu} - b^{\mu} b^{\nu}
 \ee
 is the stress-energy tensor and
 \be
 \label{eq:dual}
  F^{* \mu \nu} = b^{\mu} u^{\nu} - b^{\nu} u^{\mu}
  \ee
  is the dual of the electromagnetic field tensor given that $\rho$ is the comoving rest mass density, $p_{\rm gas}$ is the comoving gas pressure, $u^{\mu}$ is the coordinate frame 4-velocity,  $\Gamma=5/3$ is the adiabatic index of the gas, $h=1 + \Gamma/(\Gamma -1) p_{\rm gas}/\rho$ is comoving enthalpy per unit mass, $B^{i}= F^{*i0}$ is the magnetic field in the coordinate frame. Four magnetic field $b^{\mu}$ are related to 3-magnetic field $B^{i}$ as
  \bea
  \label {eq:b0}
   &&   b^0 = g_{i \mu} B^{i} u^{\mu},\\
   \label{eq:b1}
   &&  b^i  = \frac{B^i + b^0 u^i}{u^0}.
  \eea
 For diagnostics, we also use magnetic field components ($B_r, \ B_{\theta}, \ B_{\phi}$) defined in a spherical-polar like quasi-orthonormal frame as
 \be
 \label{eq:B_r_th_phi}
 B_r = B^1, \ B_{\theta} = rB^2, \ B_{\phi} = r \sin \theta B^3.
 \ee
 
 We use  the GRMHD code {\tt Athena++} \citep{White2016} to perform the simulations.
We employ the HLLE solver \citep{Einfeldt1988}  with a third-order piecewise parabolic method (PPM; \cite{Colella1984}) for spatial reconstruction. For time integration, a second-order accurate van Leer integrator  is used with the CFL number 0.3. We use a CT \citep{Gardiner2005, White2016} update  of the face-centered magnetic fields  to maintain the no magnetic monopole condition.

\subsection{Initial Condition} 
 \label{sect:method_init}

\begin{figure}
\centering
 \includegraphics[scale=0.56]{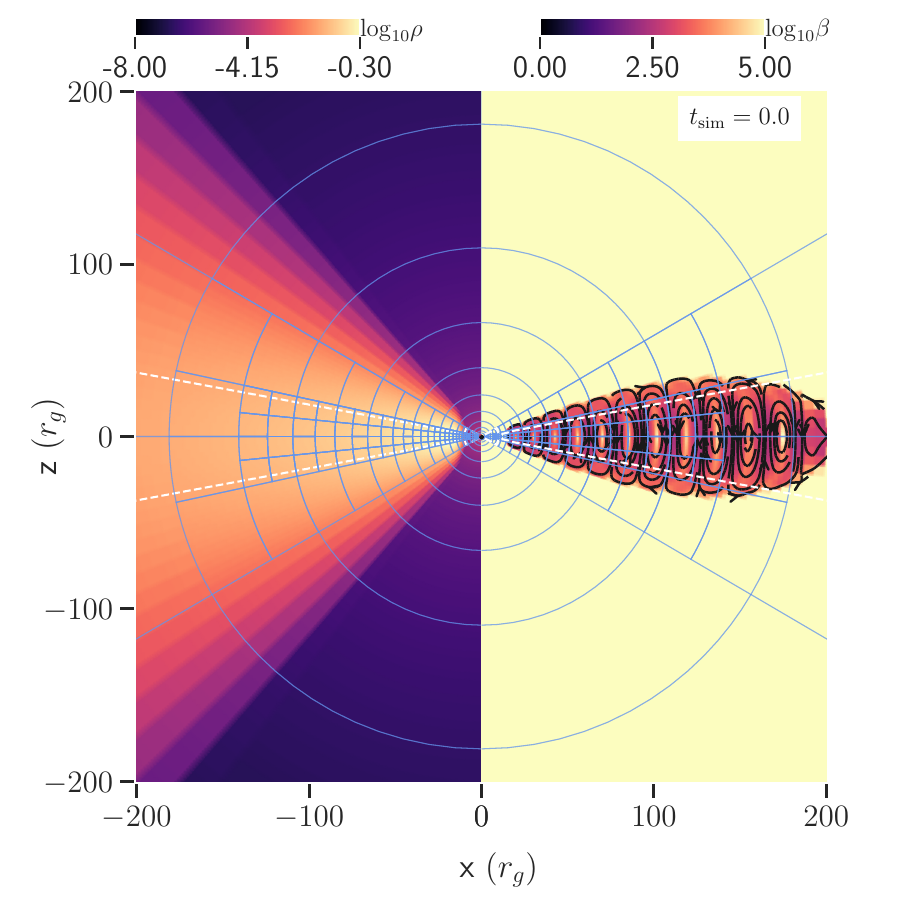}
\caption{Pictorial description of the grids and initial hydrostatic and magnetic conditions of the simulation. Each block represents a meshblock of size $16 \times 14$ in the poloidal plane in the $r$, $\theta$ directions respectively. Each meshblock has 16 grid points in the azimuthal direction. The disk aspect ratio of the initial disk is $H/R=0.23$ and is shown by the white dashed lines. Streamlines describe the initial poloidal magnetic field lines with average plasma $\beta \approx 800$.    }
\label{fig:init_grid}
\end{figure}

 We initialise a geometrically semi-thick disk of aspect ratio $\epsilon^{\text *}_{\rm in}=H_G/R=0.23$ embedded in a hot corona. Here $H_G$ is the Gaussian scale-height. The rest mass density distribution of  the initial disk is given by
 \be
    \rho_d(r,\theta) =  e^{-z^2/2H_G^2} \left( \frac{r_{\rm in}}{R} \right)^{q_d} \delta(r_{\rm in});
\ee
and the gas pressure is given by
 \be
     p_{\rm gas,d} = \rho_d c^2_{sd} = \rho_d \ \epsilon^{\text * 2}_{\rm in} \left(\frac{M_{\rm BH}}{R}\right).
 \ee
 Here, $z=r {\rm cos} \theta$, $R=r {\rm sin}\theta$ and  $\delta(r_{\rm in})=1/[1+e^{-(R-r_{\rm in})/5\epsilon^{\text *}_{\rm in} H_G}]$ is a tapping function with an inner disk radius $r_{\rm in}=15$. We consider $q_d=-1.5$ and the mass of the BH to be $M_{\rm BH}=1$. It is to be noted that the Gaussian scale height $H_G$ is related to the density-weighted scale height
 \be
 H = \frac{\int \sqrt{-g} \ \rho |\frac{\pi}{2} -\theta| \ d\theta \ d\phi } {\int \sqrt{-g} \ \rho \ d\theta \ d\phi}
\ee
as $H=\sqrt{2/\pi} H_G$ and hence, disk aspect ratio $\epsilon=\sqrt{2/\pi} \epsilon^{\text *}_{\rm in}$.

The disk is surrounded by an atmosphere defied by
\bea
    \rho_c = \rho_{\rm in} \left(\frac{r_{\rm in}}{r}\right)^{q_c}; ~~~~
 p_{\rm gas,c} = \rho_c \frac{M_{BH}}{r}
\eea
with $q_c=-1.5$ and $\rho_{\rm in}=10^{-5}$.
The tenuous atmosphere is static while the gas within the disk ($\rho \geq \rho_d$) is rotating with a Keplerian speed given by
\be
 u^{3} = \frac{r}{r-2}R^{-3/2} 
\ee 
in the Boyer-Lindquist coordinates. Also note that although two regions are set up separately and are only in approximate equilibrium,  as the system evolves and becomes MRI-active, the dynamics of the atmosphere become completely overwhelmed by the internal dynamics within the RIAF and are insensitive to the initial prescriptions in the atmosphere.
 
 In order to attain a quasi-stationary weakly magnetized RIAF (SANE; \citealt{Narayan2012}), we initialise the multiple magnetic field loops using the vector potential (\citealt{Penna2013_init})
 \begin{equation} 
 \label{eq:vec_poten}
 A_{\phi} = 
\begin{cases}
  Q~ \sin \left[ f(r) - f(r_{\rm in}) \right], ~~ Q>0 \\
  0, {\rm otherwise}.
\end{cases}
\end{equation}
 Here, 
 \bea
    && Q = C_B ~ {\rm sin}^3 \theta \left( \frac{p_1}{p_2} - p_{\rm cut} \right),\\
    && f(r) = \left(r^{2/3} + \frac{15}{8r^{2/5}} \right) \frac{1}{\lambda_B};
 \eea
 with $p_{1} (r, \theta)=p_{\rm gas} (r,\theta) - p_{\rm gas}(r_{B0},\pi/2)$,   $p_{2} (r) = p_{\rm gas} (r,\pi/2) - p_{\rm gas}(r_{B0}, \pi/2)$. The vector potential $A_{\phi}$ vanishes for $r>r_{B0}=200$.  We choose $C_B=0.5$, $p_{\rm cut}=0.4$  and $\lambda_B=0.75$  giving rise to an average plasma $\beta=\la p_{\rm gas}\ra / \la p_{\rm mag} \ra = 800$ for the initial disk (averaging is done over the region within one scale-height of the disk), $p_{\rm mag}$ being the magnetic pressure.

\subsection{Numerical setup}
\label{sect:method_setup}

We perform all the simulations of RIAFs around a non-spinning BH (spin parameter $a=0$). 
The computational domain spans over $r \in [1.94 , 300]$, $\theta \in [0, \pi]$, $\phi \in [0, 2\pi/3]$. It is to be noted that one grid point is inside the event horizon in the radial direction at the root level. This allows a causally disconnected inner boundary. Radial grids are spaced logarithmically, while meridional grids are compressed towards the mid-plane using
\be
\theta = \theta_u + \frac{1-s}{2} \ \sin 2\theta_u
\ee
with $s=0.49$ which gives rise to $\Delta \theta_{\rm pole}/\Delta \theta_{\rm eq} \approx 3.0$. Uniform grids are employed in the azimuthal direction. To improve the effective resolution, we use two levels of static refinements with a root grid resolution $160 \times 56 \times 32$ giving rise to $\Delta r : r \Delta \theta: r \Delta \phi  = 1.2:1:2.3$ at the equator in the Newtonian limit. Hence, the number of grid points increases by a  factor of two in each direction for each level of refinement. While the first level of refinement covers  $r_{L1} \in [1.94, 180], \theta_{L1} \in [\pi/3,2\pi/3]$,  the second level of refinement is applied to the region $8 < r < 140$, $4 \pi/9 < \theta < 5 \pi/9$ such that the number of $\theta$-cells per scale-height is $H/r \Delta \theta \approx 40$ in the quasi-stationary state. Hence, the effective resolution in the finest level of refinement will be $640 \times 224 \times 128$.
%See Table \ref{tab:simtab} for the details of the input parameters.

We use a pure inflow boundary condition ($u^{1} \leq 0$) at the radial inner boundary, while at the radial outer boundary, primitive variables are set according to their initial radial gradients. Magnetic fields in the inner ghost zones are copied from the nearest computation zone. On the other hand, magnetic fields at the outer ghost zones are set according to $B_r, B_{\phi} \propto r^{-2}$ while keeping $B_{\theta}$ unchanged from the last computation zone. Polar and periodic boundary conditions are used at the meridional and azimuthal boundaries, respectively.

We would also like to mention that floor values are used on different variables for numerical stability. Pressure and density are maintained throughout the simulation following
\bea
   && p_{\rm gas, floor} = {\rm max} \left(10^{-4} \  r^{-5/2}, 10^{-10}  \right), \\
   && \rho_{\rm floor} = {\rm max} \left(2 \times 10^{-4} \  r^{-3/2}, 10^{-7}  \right).
\eea 
Additionally, we also constrain the following variables as, $\beta > 0.001$, magnetization $\sigma = 2p_{\rm mag}/\rho <100$ and Lorentz factor $\gamma < 50$. It is to be noted that in the saturated state, with magnetic pressure support, the floor is applied mainly in funnels (regions close to both poles) close to the BH. This is the case for all GRMHD simulations. However, it is to be noted that the floor values do not affect the results of our RIAF simulations which are weakly magnetized.

We will introduce various diagnostics as we discuss simulation results, where the data will be averaged in different ways. Here, for future reference, we mention that the symbol `$\ \bar{} \ $' is reserved for the azimuthally averaged mean quantities, while any additional averaging (e.g. vertical or time averaging) of the quantities will be indicated by $\la . \ra$ in this paper.

\section{Evolution of the Initial RIAF Run} 
\label{sect:results_riaf}

To begin with, we would like to investigate the plausibility of the conversion of a weakly magnetized RIAF (SANE) into a highly magnetized one (MAD) due to a dynamo action. Therefore, we perform an Initial RIAF simulation as previously mentioned. We run the simulation for the time $t=1.2 \times 10^5$ to probe whether a SANE to MAD conversion occurs. In this section, we describe the evolution of the Initial RIAF towards the stationarity, its convergence and magnetic state.

\subsection{Flow evolution of the Initial  RIAF}
\label{sect:flow_evo_riaf}
\begin{figure*}
\centering
\gridline{\fig{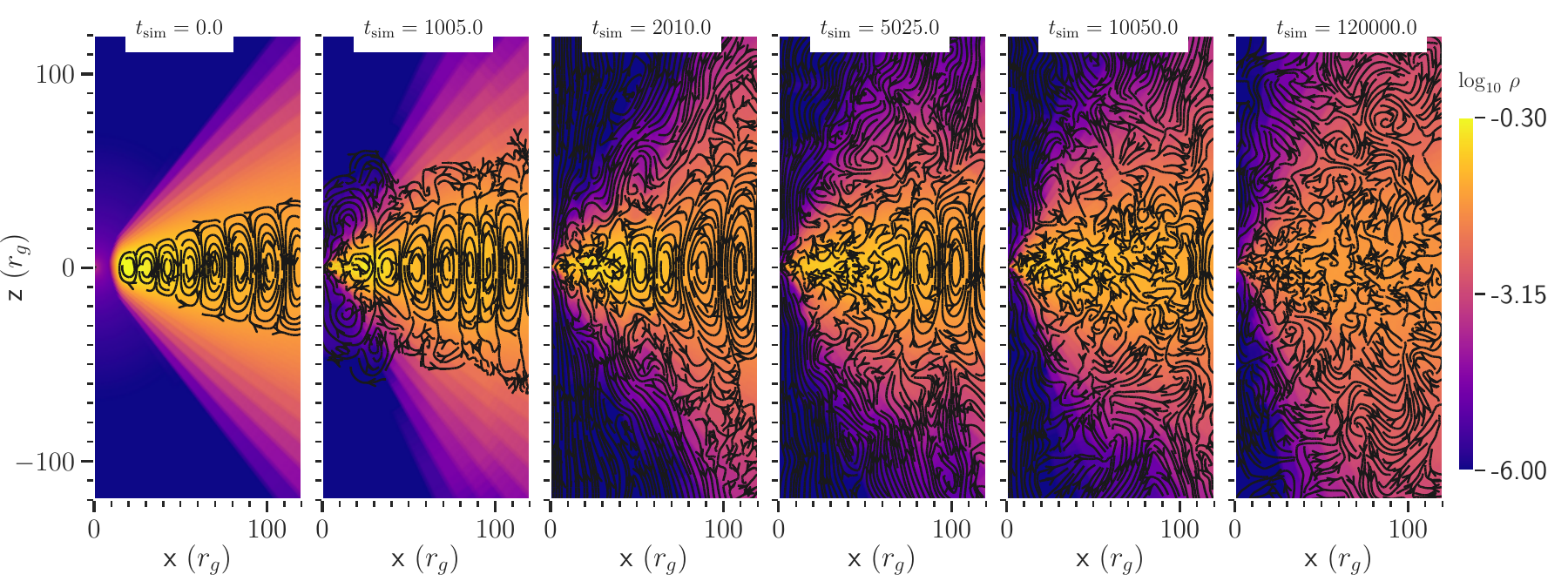}{0.95\textwidth}{(a)}
          }
\gridline{\fig{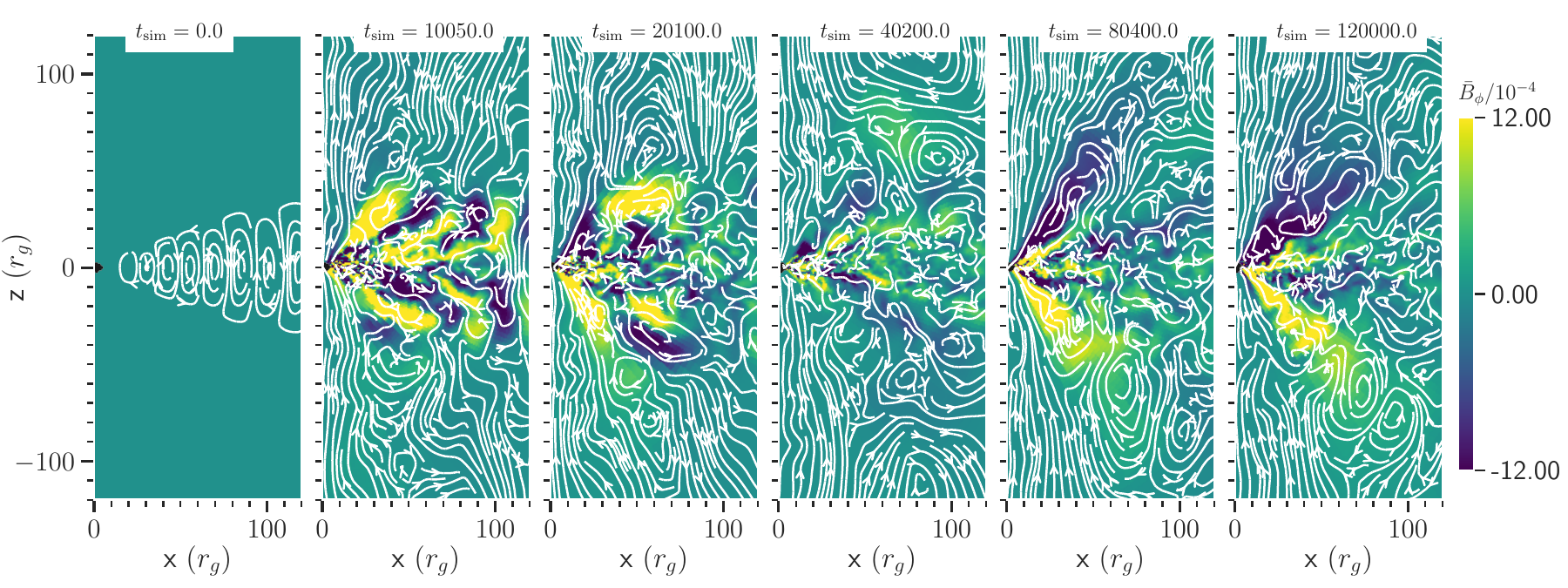}{0.95\textwidth}{(b)}
          }
\caption{Time evolution of the Initial RIAF run. Color and streamlines in the top panels show the evolution of rest mass density ($\rho$) and poloidal magnetic field  $\mathbf{B}_p=\mathbf{B}_r + \mathbf{B}_{\theta}$  respectively at an azimuthal angle $\phi=0$. The MRI grows over a few dynamical  timescales and saturates into turbulence afterwards. The bottom panels show the  evolution of the mean ($\phi$-averaged) poloidal ($\bar{\mathbf{B}}_p=\bar{\mathbf{B}}_r + \bar{\mathbf{B}}_{\theta}$) and toroidal ($\bar{\mathbf{B}}_{\phi}$) magnetic fields with time in the Initial RIAF run. The colour shows the mean toroidal field, while streamlines describe the mean poloidal field lines. Poloidal field loops of alternate polarity thread the initial disk. Shortly afterwards, the toroidal magnetic field becomes the dominant component  due to background shear. The accretion flow largely forgets the initial field configuration around $t=40200$, entering a quasi-stationary phase with an MRI-driven dynamo in action.}
\label{fig:flow_evo}
\end{figure*}

Top and bottom panels of Fig. \ref{fig:flow_evo} show the time evolution of the flow for the Initial RIAF run. We particularly focus on the initial stage of evolution of RIAF in Fig. \ref{fig:flow_evo}(a), where we show how the rest  mass density ($\rho$) and poloidal magnetic fields $\mathbf{B}_p=\mathbf{B}_r + \mathbf{B}_{\theta}$ vary in time. In contrast, Fig. \ref{fig:flow_evo}(b) focuses on the evolution of the mean ($\phi$-averaged, for definition, see equation \ref{eq:mean_B}) magnetic fields at late stages.

The first panels of Fig. \ref{fig:flow_evo} (a) and (b) show the magnetic initial condition- poloidal field loops of alternate signs with average  $\beta_{\rm av}= \la p_{\rm gas}\ra/ \la p_{\rm mag} \ra=800$, aiming to achieve a weakly magnetized RIAF (SANE; \citealt{Narayan2012}) in the quasi-stationary phase. Shear in the accretion flow converts the poloidal field into the toroidal field, while MRI amplifies the poloidal field. Therefore,  both poloidal and toroidal fields grow exponentially in a
dynamical time ($t_{\rm dyn} \approx 1/\Omega \propto R^{3/2}$). As a result, MRI grows faster in the disk close to the BH. Hence, the disk close to the BH breaks up earlier compared to that further away. After few dynamical time, the system likely enters  the non-linear regime under the influence of parasitic instabilities (\citealt{Goodman1994}) or due to different super-Alfvénic rotational instabilities (SARIs; \citealt{Goedbloed_Keppens2022}), and finally, fully MHD turbulence is developed throughout the disk.

The second panel of Fig. \ref{fig:flow_evo} shows the time $t=10050$ when MHD turbulence is fully developed throughout the region of interest ($r \leq 120$). However, it is worth noting that the system still remembers the initial field geometry as indicated by the alternate signs of mean toroidal fields at different radii. As time evolves, alternate polarity fields reconnect, and the accretion flow gradually removes the signature of initial field geometry, as can be seen in the third panel of Fig. \ref{fig:flow_evo}. Around the time $t=40200$ (fourth panel of Fig. \ref{fig:flow_evo}), the accretion flow largely forgets its magnetic initial condition and the magnetic fields generated due to an in-situ dynamo start to dominate. Finally, the subsequent disk evolution is self-regulated with a combination of MRI turbulence, dynamo and angular momentum transport. A point to be noted is that in the quasi-stationary phase, the toroidal magnetic field is always the dominant component comprising almost $85 \%$ of the total magnetic field energy. We also examined (but not shown in figures) that this ratio of toroidal to poloidal magnetic field energy modestly decreases towards the surface and within the ISCO.

\begin{figure*}
\includegraphics[scale=0.75]{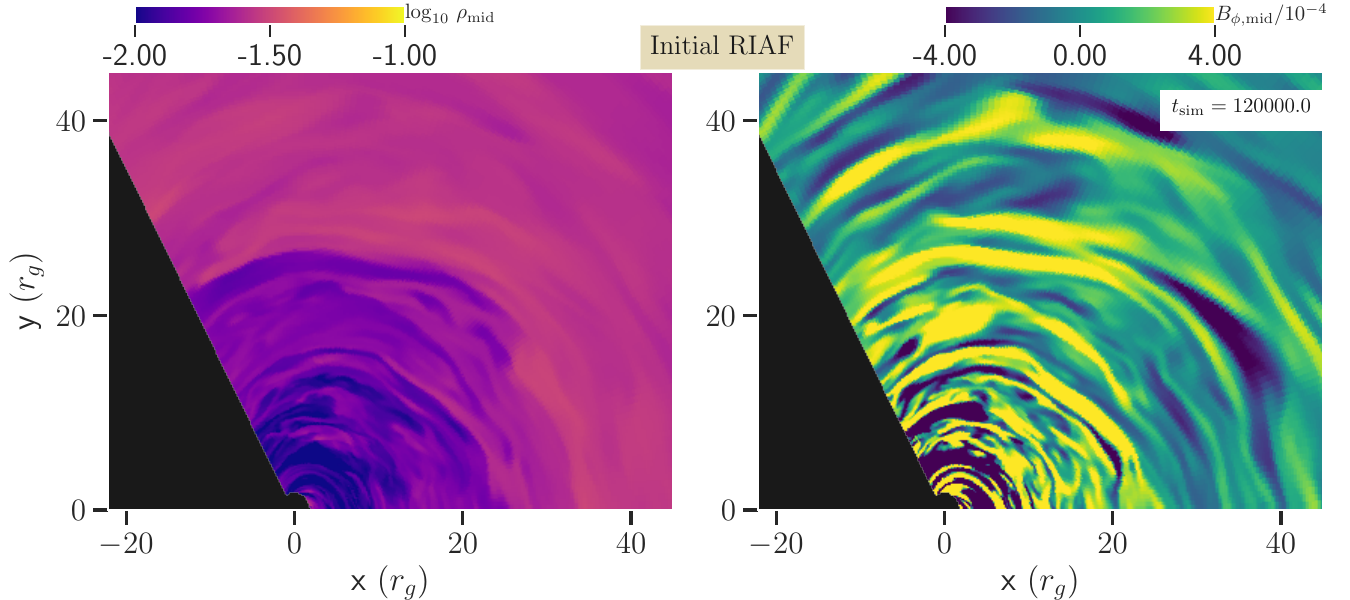}
\caption{Azimuthal structure of the rest mass density and toroidal magnetic field in the disk-midplane for the Initial RIAF run. Elongated structures  in the phi-direction are observed, which is expected in the shear-dominated accretion flow. }
\label{fig:azimuth_flow}
\end{figure*}

We showed the flow and magnetic field structures in the poloidal plane in Fig. \ref{fig:flow_evo}. Fig. \ref{fig:azimuth_flow} shows the azimuthal structures of the rest mass density (left-hand panel) and toroidal magnetic field (right-hand panel) in the disk-midplane ($\theta=90^{\circ}$) at late times for the Initial RIAF run. Both density and magnetic field show elongated structures in the $\phi$-direction, which is expected in a shear-dominated accretion disk. It is also noteworthy that the magnetic fields generated by a dynamo in the quasi-stationary phase of RIAF are not only of large scales in the radial direction (can be comprehended by the radially extended structures of mean poloidal and toroidal fields in the last panels of Fig. \ref{fig:flow_evo} (b)), 
they are also of large scales in the azimuthal direction as inferred from the snapshot of toroidal magnetic fields in Fig. \ref{fig:azimuth_flow} (also see \citep{Dhang2019}). However, the strength of the dynamo-generated large-scale field is insufficient to form a MAD, as we will discuss in the section \ref{sect:sane_or_mad}. 

\subsection{Convergence}

\label{sect:convergence}
\begin{figure}
\centering
 \includegraphics[scale=0.55]{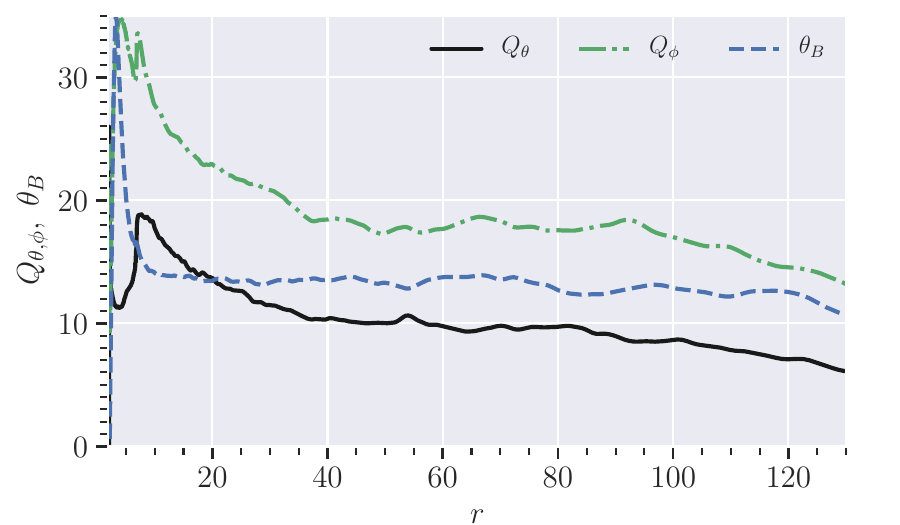}
\caption{Time averaged radial profiles of the quality factors ($Q_{\theta}$ and $Q_{\phi}$) and the magnetic tilt angle $\theta_B$ (in degrees) in the mid-plane for the Initial  RIAF run. Time average is done over $t=(5-10) \times 10^4$.  The simulation is well resolved for $r \leq120$ and resolvability starts declining afterwards. }
\label{fig:quality_fac}
\end{figure}

Before analyzing simulation results, we first verify that our simulations have achieved proper numerical convergence.  Numerical convergence implies that physically important observables (e.g. mass accretion rate) should not change significantly with the change in numerical resolution. Ideally,  we are supposed to run simulations with different resolutions and compare the results and find the minimum grid resolution required to achieve convergence. However, the GRMHD simulations we performed are computationally quite expensive. Therefore, to test convergence, we calculate different numerical metrics which were found to be useful in defining the convergence of the MRI turbulence in earlier studies (\citealt{Sorathia2012, Hawley2013}). In this work, we focus on the quality factors
\bea
&& Q_{\theta} = \frac{2 \pi}{\Omega} \frac{| \bar{b}^{\hat{\theta}} |} {\sqrt{{\overline{w}_{\rm tot}}}} \frac{1} {dx^{\hat{\theta}}}, \\
&& Q_{\phi} =  \frac{2 \pi}{\Omega} \frac{| \bar{b}^{\hat{\phi}} |} {\sqrt{{\overline{w}_{\rm tot}}}}  \frac{1} {dx^{\hat{\phi}}}
\label{eq:quality_f}
\eea
and the magnetic tilt angle 
\be
\theta_{B} = -\frac{\overline{b^{\hat{r}}b^{\hat{\phi}}}} {\overline{p}_{\rm mag}}
\label{eq:tilt_f}
\ee
measured in an orthonormal fluid frame (\citealt{White2019_tilt}). Here the angular velocity is defined as $\Omega (r,\theta)= \bar{u}^{3}/\bar{u}^{0}$, and total entropy is given by $\overline{w}_{\rm tot} (r,\theta) = \overline{\rho + \Gamma/(\Gamma-1) p_{\rm gas} + p_{\rm mag}}$. Line elements are given by $dx^{\hat{\theta}} = g_{\mu \nu} e^{\mu}_{\hat{\theta}} dx^{\nu}_{BL}$, $dx^{\hat{\phi}} = g_{\mu \nu} e^{\mu}_{\hat{\phi}} dx^{\nu}_{BL}$, where  $dx^{\mu}_{BL} = \left[0,~\Delta r,~ \Delta \theta, ~\Delta \phi -a r/(r^2 - 2Mr + a^2)  \Delta  \right]$. 
Quality factors $Q_{\theta}$ and $Q_{\phi}$ provide the information on the number of cells across a wavelength of the fastest growing mode in the $\theta$ and $\phi$- directions respectively; while $\theta_B$ measures the magnetic field anisotropy, a key factor behind angular momentum transport. 
Magnetic tilt angle above a critical value confirms the transition from linear growth of MRI to saturated turbulence (\citealt{Pessah2010}).
Earlier studies suggested that the toroidal and poloidal resolutions are coupled and the product of the quality factors $Q_{\theta} Q_{\phi} \geq 200-250$ is a good indicator for convergence in the MRI simulations (\citealt{Sorathia2012, Narayan2012, Dhang2019, grmhd_code2019}). In the meantime, we note that a unique feature of the magnetic tilt angle is that it does not change with an increasing resolution for converged simulations; $\theta_B$ shows a narrow range of value $10^{\circ}-14^{\circ}$ for the converged runs (e.g \citealt{Sorathia2012, Hogg2018a, Dhang2019}) and turned out to be a better indicator of convergence.
 
The top and bottom panels of Fig. \ref{fig:quality_fac} shows the radial profiles of the average (averaged over $\phi$, $\theta$ and time) quality factors ($\la Q_{\theta} \ra$, $\la Q_{\phi} \ra$),  and magnetic tilt angle ($\la \theta_B \ra$) close to the mid-plane of the disk for the Initial RIAF run. Meridional average is done over one scale-height (with $H/R=0.2$) above and below the mid-plane, while azimuthal average is done over all cells. The time average is done over the  time interval $t=(5-10) \times 10^4$. While the quality factors indicate that our simulation is marginally resolved with $\la Q_{\theta} \ra \la Q_{\phi} \ra \gtrsim 180$ up to $r=100$, the radial profile of $\la \theta_B \ra $ clearly shows that our Initial RIAF simulation is well resolved till $r=120$ and resolvability starts to decline afterwards because of the poor resolution at larger radii.

\subsection{Characterizing the magnetic state of the Initial RIAF?}
\label{sect:sane_or_mad}

\begin{figure}
\centering 
 \includegraphics[scale=0.75]{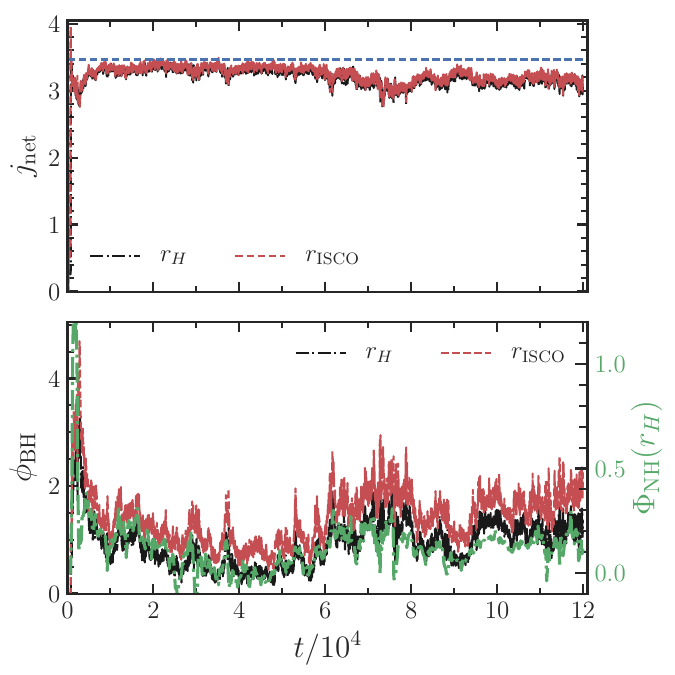}
\caption{The time history of accreted specific angular momentum $j_{\rm net}$ and MAD parameter $\phi_{\rm BH}$ calculated at the event horizon and at the ISCO for the Initial  RIAF run. Blue dashed horizontal line in the top panel  corresponds to the Keplerian value of specific angular momentum at the ISCO. For correspondence, time variation of the magnetic flux $\Phi_{\rm NH}$ (equation \ref{eq:flux_r}) threading the BH is also shown.}
\label{fig:jnet_mad_param}
\end{figure}

In this section, we characterize our Initial RIAF simulation, particularly examining the indicators that distinguish the SANE from the MAD state. Following, \cite{Narayan2012}, we study the time evolution of the specific angular momentum of the accreting material
 \be
 j_{\rm net}(r,t) = -\frac{1}{\dot{m}}\int T^{1}_{3}  \ dS_r
 \ee
 and the MAD parameter 
 \be
\phi_{BH}(r,t) = \frac{\sqrt{4 \pi}}{2\sqrt{\dot{m}}}\int |B^1|  \ dS_r
\ee
to investigate the magnetic state of the Initial RIAF run. Here, the mass accretion rate at any radius $r$ is defined as
\be
\dot{m} (r,t) = - \int  \rho u^{1} \ dS_r,
\ee
where area element is given by $dS_r = \sqrt{-g} \ d\theta \ d\phi$, and  the integration is performed over all $\theta$ and $\phi$.
 
The MAD parameter $\phi_{BH}$ is a dimensionless number which is found to be useful in characterizing the magnetic state of the simulations. Earlier studies suggest that an accretion flow attains a MAD state once $\phi_{BH}$ reaches a critical value $\phi_{BH,c}\approx 40$ at the event horizon (\citealt{Tchekhovskoy2011}). Additionally, $j_{\rm net}$ also shows a highly sub-Keplerian nature at the event horizon for MAD simulations, where angular momentum transport is highly efficient due to the large-scale Maxwell stress. On the contrary, $j_{\rm net}$ maintains a slightly sub-Keplerian value at the event horizon in the SANE simulations (\citealt{Narayan2012}). Moreover, $j_{\rm net}$  is shown to be a good indicator of convergence in the MRI-active turbulent accretion flow. For a converged simulation, $j_{\rm net}$ maintains a sub-Keplerian value inside the ISCO with a non-decreasing trend in time throughout the simulation (\citealt{Hawley2013, Dhang2019}). 
 
Fig. \ref{fig:jnet_mad_param} show the time evolution of $j_{\rm net}$ and $\phi_{BH}$ at two different radii, at the ISCO  and at the event horizon. We also plot the time variation of signed flux threading the event horizon ($r_H=2$) of the BH in the northern hemisphere, $\Phi_{NH} (r_H)$ (equation \ref{eq:flux_r}) for a future reference in section \ref{sect:results_flux}. The  value of $\phi_{\rm BH}$ always remains around one which is well below the value  ($\ge 40$) required for the MAD state. Such a low value of $\phi_{\rm BH}$ implies that the magnetic state of the Initial RIAF run is in the SANE regime. Slightly sub-Keplerian value of specific angular momentum  $j_{\rm net}$ is also an indicator of the SANE magnetic state of our initial RIAF simulation.

\subsection{Inflow equilibrium}

\begin{figure}
\centering
 \includegraphics[scale=0.55]{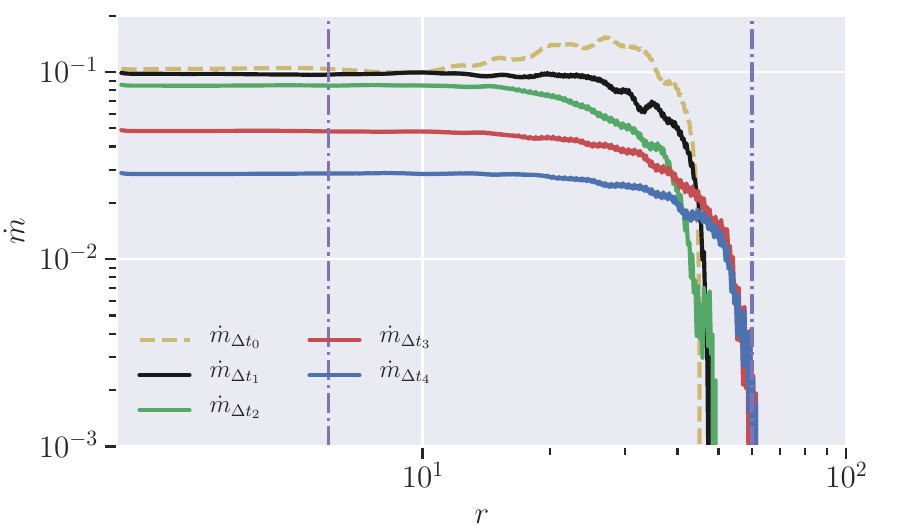}
 \includegraphics[scale=0.54]{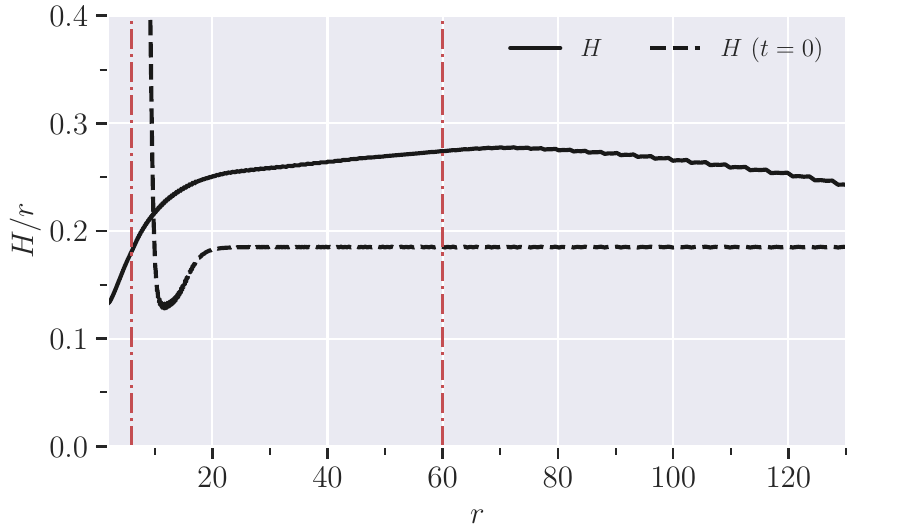}
\caption{Top panel: Time and spatial ($\theta$ and $\phi$) averaged accretion rates as a function of radius for the Initial RIAF run. Time averages are done over five different intervals $\Delta t_0=(3.75-7.5)\times 10^3$,  $\Delta t_1=(7.5-15)\times 10^3$, $\Delta t_2=(1.5-3)\times 10^4$, $\Delta t_3=(3-6)\times 10^4$ and $\Delta t_4=(6-12)\times 10^4$.  It can roughly be inferred that the inflow equilibrium radius is $r_{\rm eq} = 60$  for the  Initial RIAF run.\\ Bottom panel: Disk aspect ratios $\epsilon=H/r$ for the Initial RIAF  run in the quasi-stationary state.  Simultaneously, we also plot its initial radial profile to study the change over time. Time averages are done over $\Delta t_4=(6-12)\times 10^4$. }
\label{fig:mdot_H_riaf}
\end{figure}

We will inject external magnetic flux in the quasi-stationary turbulent RIAF to study magnetic flux transport (ref. section \ref{sect:results_flux}). Therefore, it is important to find out the inflow equilibrium radius - the radius within which flow attains a quasi-stationary state, for the Initial RIAF run.  
Following \cite{Narayan2012}, we investigate the variation of average mass accretion rate $\la \dot{m}(r) \ra$ with time to find out the
inflow equilibrium radius. Spatial averages are done over all $\theta$ and $\phi$. We use five different intervals $\Delta t_0=(3.75-7.5)\times 10^3$,  $\Delta t_1=(7.5-15)\times 10^3$, $\Delta t_2=(1.5-3)\times 10^4$, $\Delta t_3=(3-6)\times 10^4$ and $\Delta t_4=(6-12)\times 10^4$ to do the time average. The top panel of Fig. \ref{fig:mdot_H_riaf} shows $\la \dot{m}(r) \ra$ at different time intervals for the Initial RIAF run. It can be inferred from the radial profiles of $\dot{m}$ that the inflow equilibrium radius for the Initial RIAF run reaches $r_{\rm eq}\approx60$ at late times. This is the radius that guides us to determine the injection radius for the external magnetic field loops.

The bottom panel of Fig. \ref{fig:mdot_H_riaf} shows the radial variation of disk aspect ratio  $\epsilon=H/r$  in the quasi-stationary state. Although, overall, the disk aspect ratio epsilon fluctuates over time at $\lesssim 20\%$ level, when averaged over $\Delta t_4$, the disk aspect ratio $\epsilon$ slowly increases with the increasing radius until the inflow equilibrium radius and its value lie around $\epsilon = 0.25$ for $r\gtrsim20$, where general relativistic effects are negligible. Such a variation of scale height in our simulation is also in agreement with that observed in previous GRMHD simulations of the SANE RIAF (e.g \citealt{Narayan2012}).

\subsection{Large-scale magnetic field and dynamo}

\begin{figure}
\centering
 \includegraphics[scale=0.55]{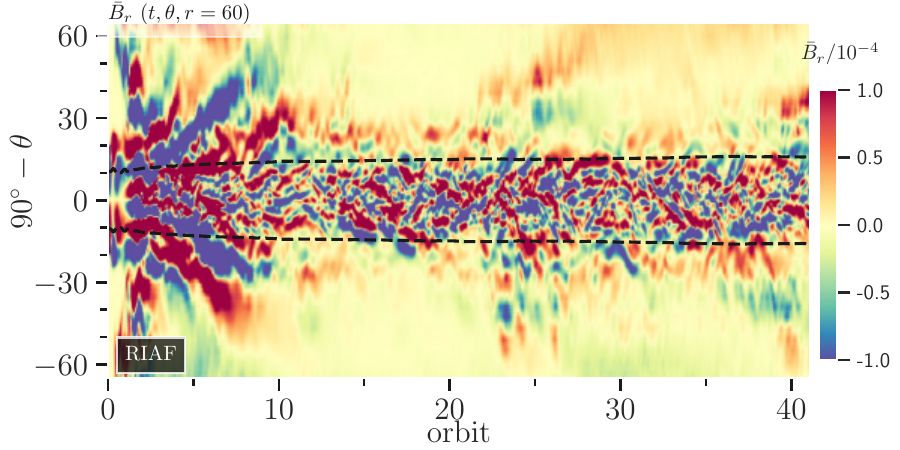}
 \includegraphics[scale=0.55]{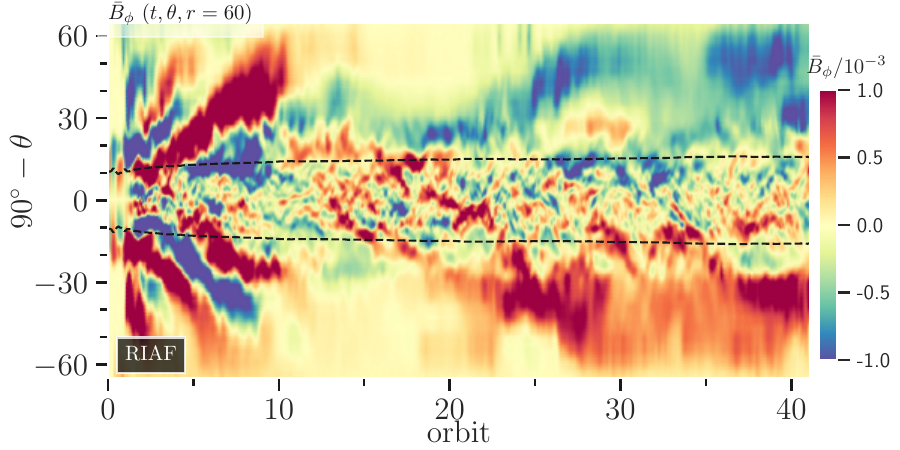}
\caption{The butterfly diagram: Space-time plots of mean radial $B_r (r=60,, \theta,t)$ (top panel) and toroidal  $\bar{B}_{\phi}(r=60,\theta,t)$ (bottom panel) for the Initial RIAF run. Time is expressed in units of local orbit at $r=60$. Both radial and toroidal fields show irregularity, which is typical feature of dynamo in a geometrically thick RIAF. }
\label{fig:butter_br_bphi}
\end{figure}

We find that our Initial RIAF simulation is in the SANE state and an MRI dynamo generates the large-scale magnetic fields and governs the magnetic field evolution at late times as discussed in section \ref{sect:flow_evo_riaf}. To characterize the dynamo action, it is customary to visualise the spatio-temporal variation of the mean magnetic field to investigate dynamo. We define the mean magnetic 
field as the azimuthally averaged field
\be
\label{eq:mean_B}
\bar{B}_{i} (r,\theta) = \frac{1}{\phi_{\rm ext}} \int_{0}^{\phi_{\rm ext}} B_i(r,\theta,\phi) d \phi,
\ee 
where $\phi_{\rm ext} $ is the extension in the $\phi$ direction, and $i \in (r, \theta, \phi)$.  Fig. \ref{fig:butter_br_bphi} shows the variation of mean radial $B_r (R_0,, \theta,t)$ (top panel) and mean toroidal field $\bar{B}_{\phi}(R_0,\theta,t)$ (bottom panel) with latitude ($90^{\circ} -\theta$) and time at a radius $R_0=60$. This is also known as the butterfly diagram. Both radial and toroidal fields show irregular behaviour in their butterfly diagrams. Additionally, the radial field is less coherent compared to the toroidal field as observed in earlier studies of the dynamo in the SANE RIAF \citep{Hogg2018a, Dhang2020}.  This intermittent dynamo cycle in the RIAF is in  contrast to the very regular dynamo cycles observed in a thin Keplerian disk (e.g. see \cite{Flock2012a}). Irregularity in the dynamo cycle arises because of the slightly sub-Keplerian angular velocity of the geometrical thick RIAF (\citealt{Dhang2019}).  

Earlier studies found that while a large-scale dynamo generates large-scale magnetic fields in the high latitudes, a fluctuation dynamo dominates close to the disk mid-plane suppressing the production of the large-scale magnetic field there (\citealt{Dhang2019}). This can be qualitatively understood by looking at the large and coherent magnetic structures (especially for the toroidal fields) in the high latitudes, while more patchy distribution near the disk mid-plane ($90^{\circ}-\theta=0^{\circ}$) in the butterfly diagram in Fig. \ref{fig:butter_br_bphi} and also in the last two panels of Fig. \ref{fig:flow_evo}. However it should be emphasized that although the MRI dynamo does produce a large-scale magnetic field, it is not strong enough to create a MAD, which is conducive for strong jets (section \ref{sect:sane_or_mad}). This inefficiency is likely to be due to the weak $\alpha$-effect (\citealt{Dhang2020}) which is responsible for poloidal field generation. Additionally, a strong turbulent pumping present in MRI-active RIAF tends to prevent accumulation of large-scale magnetic field near the BH as suggested in \cite{Dhang2020}.

\section{Transport of external magnetic field loops}
\label{sect:results_flux}
\begin{figure*}
\centering
 \includegraphics[scale=0.55]{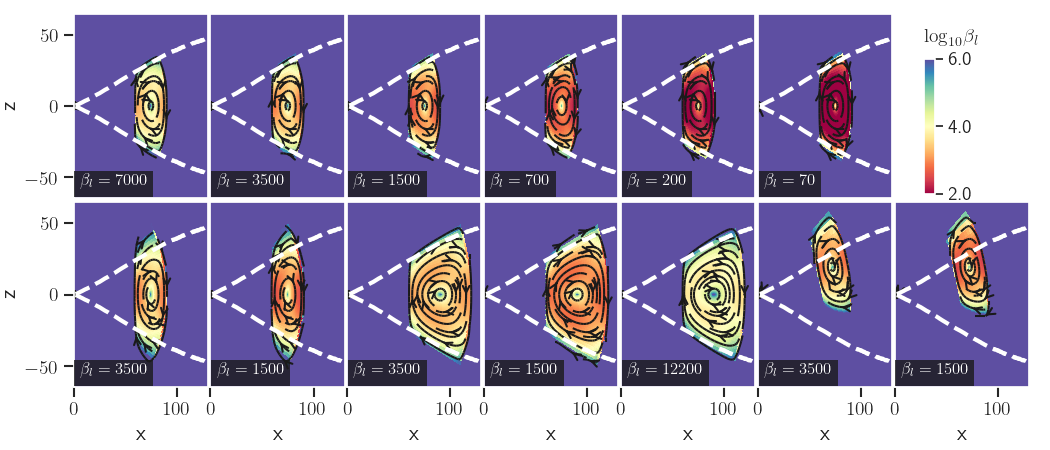}
\caption{Different types of external field loops injected at $t=8.06 \times 10^4$.  White dashed lines mark $z=1.5H$ above and below the mid-plane. For details, see Table \ref{tab:loop}.}
\label{fig:beta_loop}
\end{figure*}

In this section, we study the accretion of external magnetic flux injected on top of the fully turbulent SANE state obtained in Section \ref{sect:results_riaf}. Our aim is to investigate whether or not the system can bring in external magnetic flux available at the outer radii, all the way to the central BH that may eventually lead to a MAD state.
While the actual configuration of the external field is unknown and could be complex, we anticipate it is likely to be closed. Therefore, instead of the commonly used net vertical field, we inject the poloidal magnetic field loops of different strengths, and radial and vertical sizes as shown in Fig. \ref{fig:beta_loop} and study their transport. As the controlled experiments, these field loops confined between the radii $r_{l_1}$ and $r_{l_2}$ are prescribed by
\be
\label{eq:loop}
A_{\phi, l} = \sqrt{\frac{2 p_{\rm gas}(r_{lc},\pi/2)}{C_{l}}}\left[\frac{\rho(r,\theta^{\prime})}{\rho(r,\pi/2)} - \delta_{l}\right]^2  \sin \left[ \kappa(R-r_{l1})\right] 
\ee
where $p_{\rm gas}$, $\rho$ are the initial pressure and density profiles respectively, $\theta^{\prime}=\theta + \theta_{\rm shift}$, $\kappa = \pi/(r_{l_2}-r_{l_1})$ and $r_{lc}=(r_{l_1}+r_{l_2})/2$. Vanishing $\theta_{\rm shift}$ implies that the loop centre is 
at the mid-plane, while a positive value of $\theta_{\rm shift}$ indicates that the loop is off-centred.
The vertical size of the loop is set by $\delta_l$. The magnetization of the loop is controlled by the parameter $C_l$ and characterized by $\beta_l=\la p_{\rm gas} \ra/ \la p_{\rm mag,l} \ra$, where $\la p_{\rm gas} \ra$ and $\la p_{\rm mag,l} \ra$ are the gas pressure of the Initial RIAF at the time of loop injection and magnetic pressure of the injected loop respectively. Additionally, note that the average is performed within the loop. We choose $r_{l_1}$ to be the inflow equilibrium radius $r_{eq}=60$, and different values of $r_{l_2}$as tabulated in Table \ref{tab:loop}. 

We restart the Initial RIAF run at $t= 8.06 \times 10^4$, inject the external field loops (equation \ref{eq:loop}) and run till $t=t_{\rm end}$ as tabulated in Table \ref{tab:loop}. It is to be noted that we also run the Initial RIAF simulation longer to compare it with the simulations with injected magnetic field loops.
%We considered plasma $\beta_l$,  (which is proportional to $C_l$, see Table \ref{tab:loop}), as the fiducial parameter for the injected loops. 
We injected loops of different strengths with a wide range of plasma $\beta$ ranging from $\beta_l=7000$ (weak but stronger than the pre-existing mean fields produced by MRI dynamo) to $\beta_l=70$ (very strong field typically used in MAD simulations, but of much larger size than that used in our simulations). We also explored the effects of other parameters such as radial, vertical sizes and injection latitude of the loops on the flux transport process while considering the loops of fiducial plasma $\beta$ values $\beta_l=3500$ and $\beta_l=1500$ respectively. Additionally, we studied the transport of  large magnetic loops of strength (of $\beta_l=12200$) similar to that of the mean fields produced by MRI dynamo in the quasi-stationary phase of the Initial RIAF run. The configurations of injected field loops from all these restarts are illustrated in Figure \ref{fig:beta_loop}.

\begin{deluxetable*}{cccccccccc}
\tablenum{1}
\tablecaption{Details of the injected external field loops characterized by plasma $\beta_l$, vertical size $z_l$ and confinement radii between $r_{l1}$ and $r_{l2} \ge r_{l1}$. $\theta_{\rm shift}$ and $\Phi_{l, {\rm max} }$ are the tilt of the loop with respect to the mid-plane and the total magnetic flux in the injected loop in code units, respectively. Magnetic field loops are injected at $t=8.06 \times 10^4$ and run till $t=t_{\rm end}$.  \label{tab:loop}}
%\centering
%\begin{tabular}{lp{2.5cm}lp{1.8cm} p{1cm} p{1cm} p{1cm} p{1cm} p{1cm} p{1cm} p{1cm} p{1cm} }
 %\hline
 %\multicolumn{10}{|c|}{Parameters} \\
 %\hline
 \tablehead{
  \colhead{Name} & \colhead{$C_l$}   & \colhead{$\delta_l$}   & \colhead{$r_{l1}$}	& \colhead{$r_{l2}$} & \colhead{$z_{l} $} & \colhead{$\beta_{l}$} & \colhead{$\theta_{\rm shift}$} & \colhead{$\Phi_{l, {\rm max}}$} & \colhead{$t_{\rm end}/10^5$} 
 }
\decimalcolnumbers
\startdata
 Initial RIAF &- &- &- &- &- &- &- &-  & 1.2 \\
  $\beta \_7000$ & $10^{-3}$	 &0.2	  & 60	  &90         & 1.5 H      &  7000      &$0^{\circ}$ &       1.48  & 1  \\
$\beta \_3500$ & $5 \times 10^{-4}$ &0.2   &60   &90      & 1.5 H     & 3500   &$0^{\circ}$   & 2.01    & 1.2    \\
$\beta \_1500$  & $ 2.21 \times 10^{-4}$ & 0.2 &60  &90  & 1.5 H  & 1500  &$0^{\circ}$ & 2.93  & 1.2\\
 $\beta \_700$ & $10^{-4}$   &0.2  &60    &90  &1.5 H & 700 &$0^{\circ}$ & 4.28  & 1.2\\
 $\beta \_200$ & $2.78 \times 10^{-5}$ &0.2 &60 &90 &1.5H & 200 &$0^{\circ}$ &7.95  & 1\\
 $\beta \_70$ &$10^{-5}$  &0.2  &60   &90  & 1.5 H   & 70 &$0^{\circ}$ & 13.14 & 1.2\\
 $\beta \_3500 \_{\rm tall}$ & $1.168\times 10^{-3}$ &0.0016 &60 &90 &2.5H &3500  &$0^{\circ}$ & 2.04 & 1\\
$\beta \_1500 \_{\rm tall}$ & $5 \times 10^{-4}$  &0.0016  &60 &90   &2.5 H  &1500  &$0^{\circ}$ &2.97 & 1 \\
$\beta \_3500 \_{\rm big}$ &$1.429 \times 10^{-5}$ &0.2 &60 &120 &1.5H &3500 &$0^{\circ}$ &3.04 & 1\\
 $\beta \_1500 \_{\rm big}$ &$ 6.12 \times 10^{-5}$ & 0.2 &60 &120 &1.5 H & 1500  &$0^{\circ}$ & 4.42 & 1\\
 $\beta \_12200 \_{\rm big}$ & $5 \times 10^{-4}$  &0.2  &60  &120  &1.5 H  & 12200 &$0^{\circ}$ &1.70 & 1.1\\
 $\beta \_3500 \_{\rm offc}$ & $5.88 \times 10^{-4}$ &0.2 &60 &90 &1.5H & 3500 & $15^{\circ}$  & 1.56 & 1\\
  $\beta \_1500 \_{\rm offc}$ & $2.55 \times 10^{-4}$ &0.2 &60 &90 &1.5H & 1500 & $15^{\circ}$ & 2.32 & 1\\
 \enddata

\end{deluxetable*}

%\subsection{Results}
\subsection{Diagnostics}
Before discussing the results in detail, we define the following quantities used to discuss the transport 
of external magnetic flux, including:

The radial magnetic flux threading the $r=const$ surface in the northern hemisphere 
\be
\label{eq:flux_r}
\Phi_{NH} (r) = \int_{\theta=0}^{\pi/2} \int_{0}^{\phi_{ext}} \sqrt{4 \pi} B_r(r,\theta,\phi) \sqrt{-g} \ d\theta d\phi\ ,
\ee
the vertical flux threading the mid-plane region
\be
\label{eq:flux_th}
\Phi_{\rm mid} (r) = \int_{r=r_H}^{r} \int_{0}^{\phi_{ext}} - \frac{\sqrt{4 \pi}}{r} B_{\theta}(r,\theta=\pi/2,\phi) \sqrt{-g} \ dr \ d\phi\ ,
\ee
and the total flux available for accretion at different $r$ in the northern hemisphere
\be
\label{eq:flux_mid}
\Phi_{\rm tot} (r) =  \Phi_{\rm NH} (r_H) + \Phi_{\rm mid} (r);
\ee
where $r_H=2r_g$ is the event horizon radius of the BH.
We also define a normalised flux representing the efficiency of the flux transport and  defined by
\be
\label{eq:efficiency}
f_B=\frac{\Phi_{NH} (r_H)}{ \Phi_{l, {\rm max}} }.
\ee
Here, $\Phi_{l, {\rm max}}$ is the total flux at the loop centre at the time of injection or in the beginning of the simulation (only for the Initial RIAF run; also see Table \ref{tab:loop}). 

\subsection{Results for the fiducial parameter- plasma $\beta$}

\begin{figure*}

\gridline{\fig{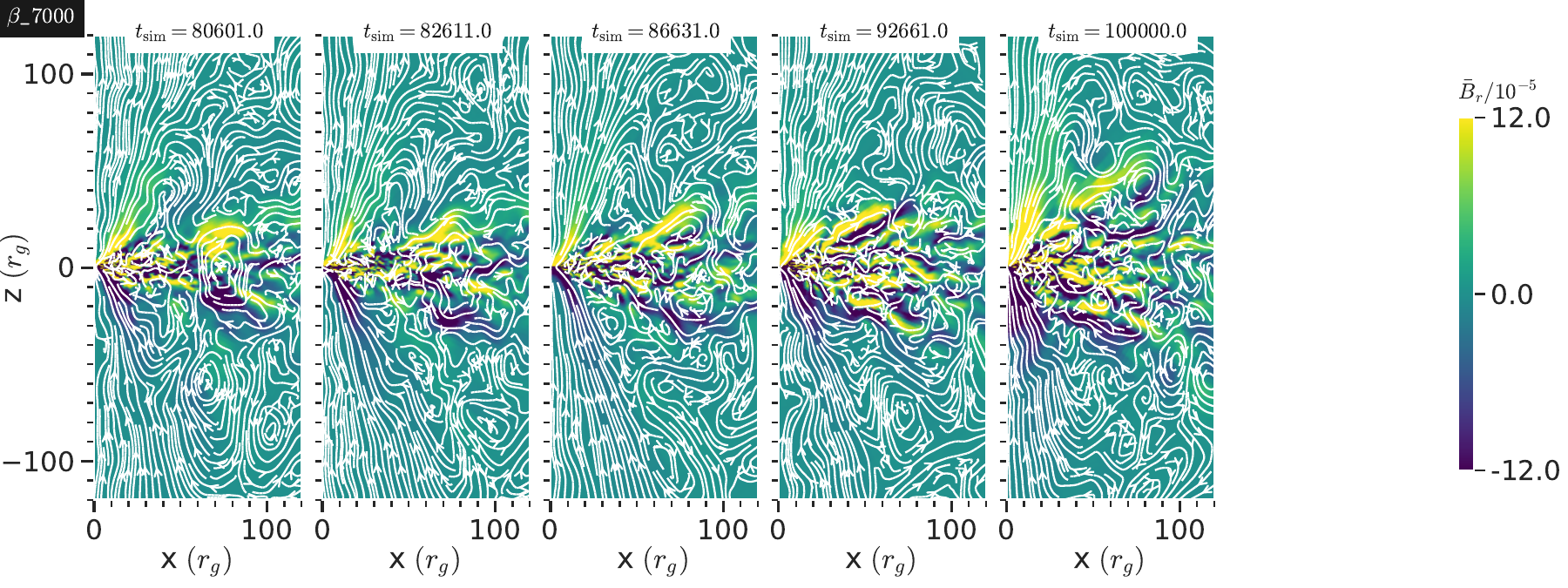}{0.95\textwidth}{(a)}
          }
\gridline{\fig{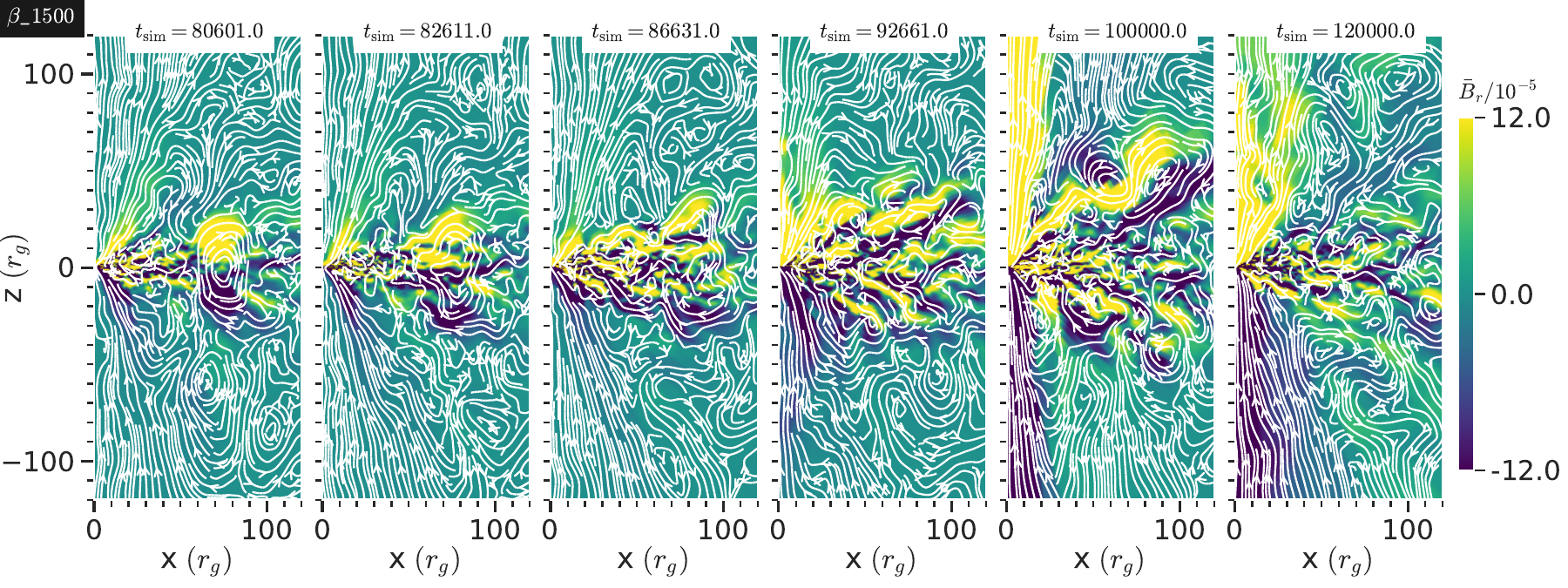}{0.95\textwidth}{(b)}
          }
\gridline{\fig{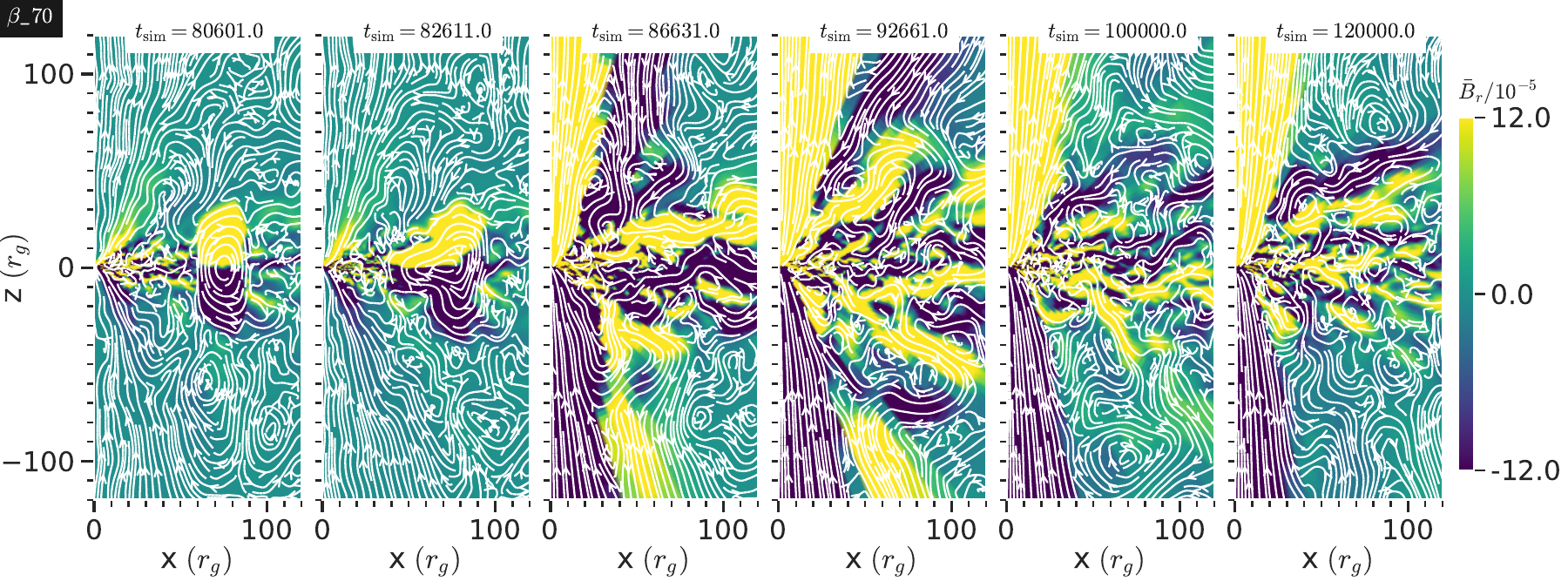}{0.95\textwidth}{(c)}
          }
\caption{Evolution of external magnetic field loops injected in a turbulent quasi-stationary RIAF. Panels at the top, middle, and bottom show the evolution of the magnetic field loops of plasma $\beta=7000$ (weak field strength), $\beta=1500$ (moderate field strength ) and $\beta=70$ (strong field), respectively. Colour represents the mean radial field $\bar{B}_r$; while poloidal field distribution $\mathbf{\bar{B}_p}=\mathbf{\bar{B}_r} + \mathbf{\bar{B}_{\theta}}$ is described by streamlines. Note that a fraction of the injected magnetic flux reaches the BH.}
%stronger the injected field loops are, shorter the accretion time. }
\label{fig:loop_evo}
\end{figure*}

First, we will discuss the dependence of the flux transport and the emergent accretion properties on the strength of the loops, defined by plasma $\beta_l$ (also see Table \ref{tab:loop} and Fig. \ref{fig:beta_loop}). We start by discussing the qualitative picture on the evolution of the magnetic flux injected between the radii $r=60$ and $r=90$ on top of the existing magnetic field in the quasi-stationary Initial RIAF. It is worth noting that the plasma $\beta$ of the total (mean + fluctuation) magnetic field is $\beta_{\rm tot}=70$, while that of the mean field alone is $\beta_{\rm mean}=12250$, for the Initial RIAF run. Fig. \ref{fig:loop_evo} shows the evolution of external magnetic field loops of three different  $\beta_l$-s. Colour shows the intensity of mean radial field $\bar{B}_r$, while streamlines describe the mean poloidal fields $\mathbf{\bar{B}_p}=\mathbf{\bar{B}_r} + \mathbf{\bar{B}_{\theta}}$.

The top panels show the time
evolution of the weakly magnetized loop of strength  $\beta_l=7000$. Injection of the weak external magnetic field loops re-excites the MRI in the accretion flow, enhances accretion stresses (see Fig. \ref{fig:acc_stress}) and hence lead to higher mass accretion rates (see Fig. \ref{fig:mdot_strength}). Poloidal flux slowly drifts towards the BH and a fraction of the injected flux accumulates near the BH (which is quantitatively shown in Fig. \ref{fig:flux_r_strength}). We see an increase in the radial magnetic flux threading the BH when compared to that in the Initial RIAF run. This can be comprehended by comparing the snapshots at $t=80601$ (the flux level close to the BH does not change significantly in the quasi-steady state of Initial RIAF as shown in Fig. \ref{fig:flow_evo}) and the last panel at $t=10^5$.

Next, we discuss the transport  of moderately strong magnetic field loops of $\beta_l=1500$ as shown in the middle panels of Fig. \ref{fig:loop_evo}. Magnetic flux reaches the BH in a shorter time compared to that in the weak field case of $\beta_l=7000$. This is due to the stronger accretion stresses produced (see Fig. \ref{fig:acc_stress}) in the accretion flow due to the injection of stronger magnetic field loops. The radial magnetic field strength in the polar region is found to be stronger than the weak field case at late times. This is because of the larger amount of flux associated with the loop of $\beta_l=1500$ than that with the loop of $\beta_l=7000$. 

Finally, we examine the transport of the strong field loops with $\beta_l=70$, similar to the strength of the total (mean + fluctuation) magnetic field in the quasi-stationary phase of Initial RIAF. Unlike the previous two weak-field cases, here the injected loops are so strong with the most unstable wavelengths comparable to the disk scale height, and that drives strong channel flows over the entire vertical extent of the disk.
%Also injection of strong field loops in the accretion flow generates 
This further generates a spike in the large-scale Maxwell stress (see Fig. \ref{fig:acc_stress}) producing a strong inflow of mass and magnetic flux. Magnetic flux reaches the BH very quickly and fills the polar region. The system remains to be in a strongly turbulent state till the end of the simulation.

\begin{figure*}
\gridline{\fig{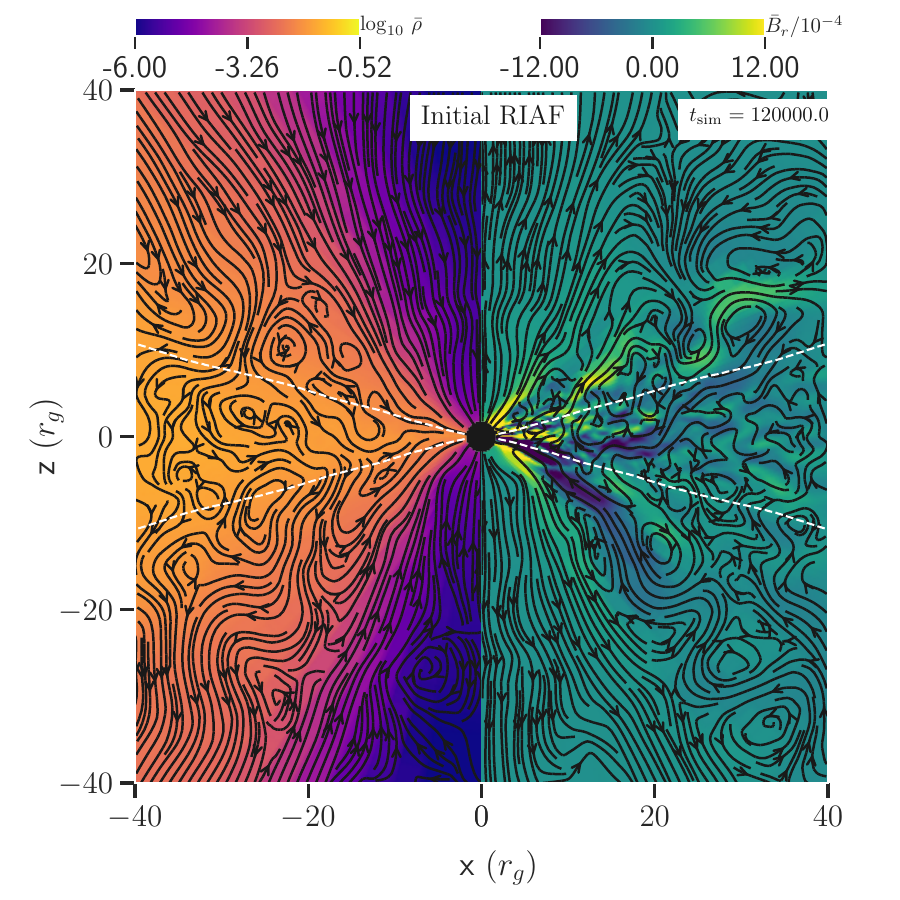}{0.49\textwidth}{(a)}
          \fig{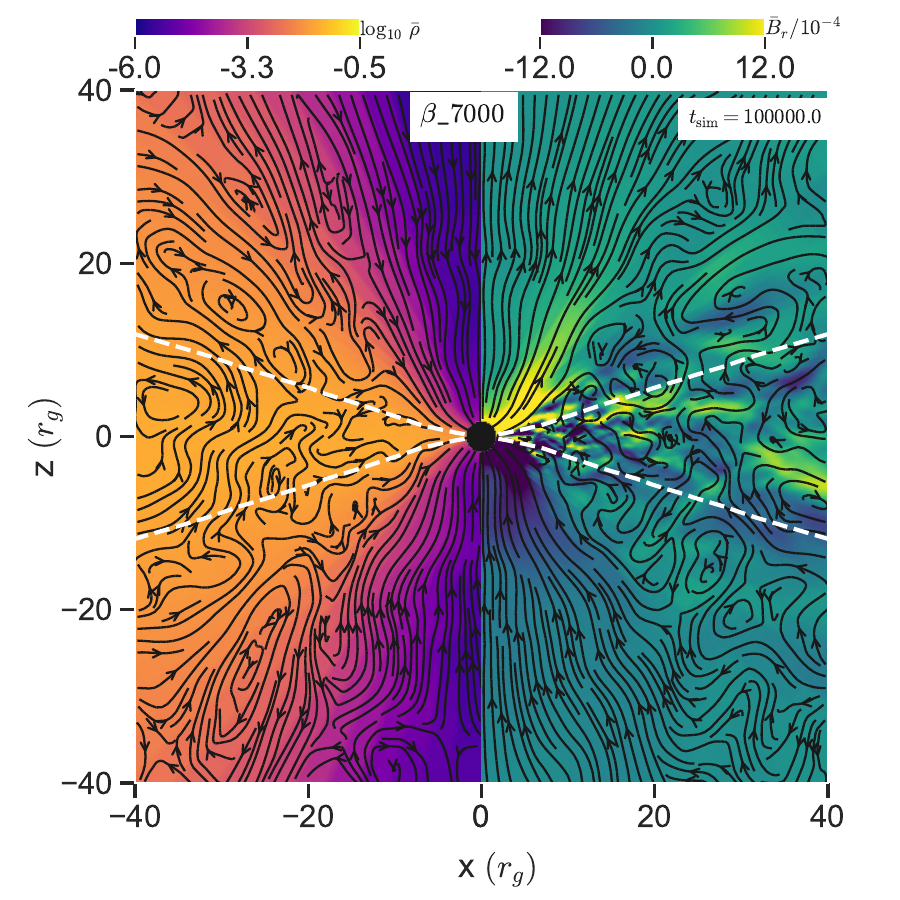}{0.49\textwidth}{(b)}
          }
\gridline{
          \fig{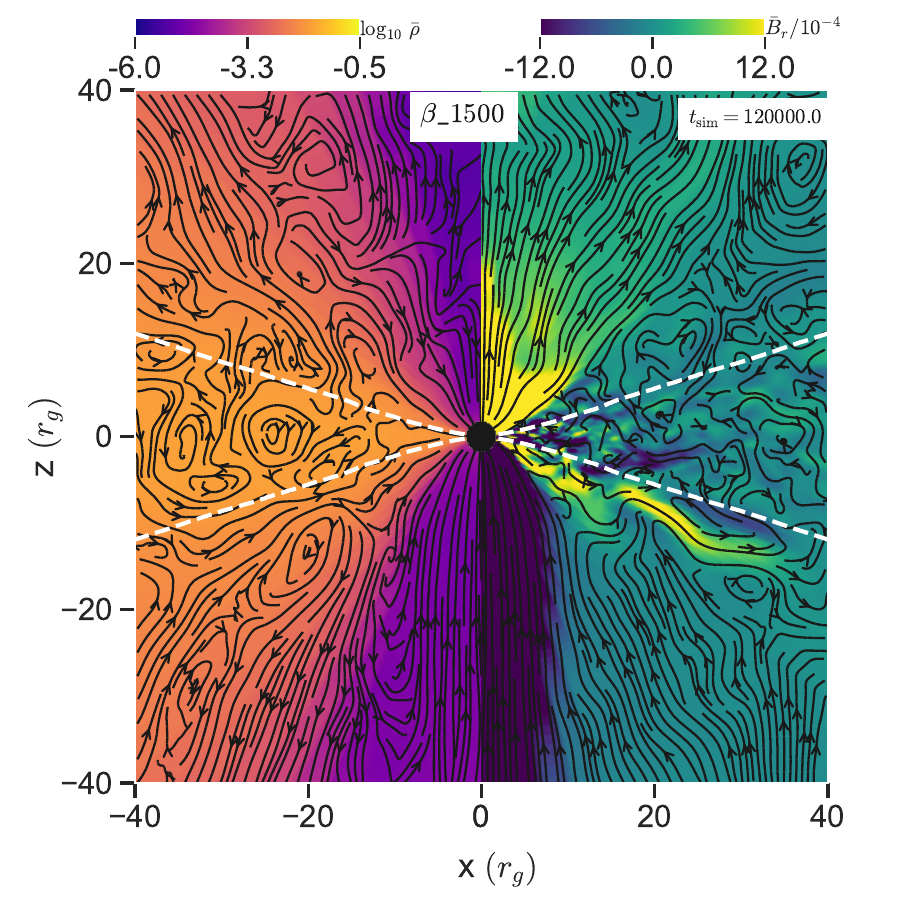}{0.49\textwidth}{(c)}
          \fig{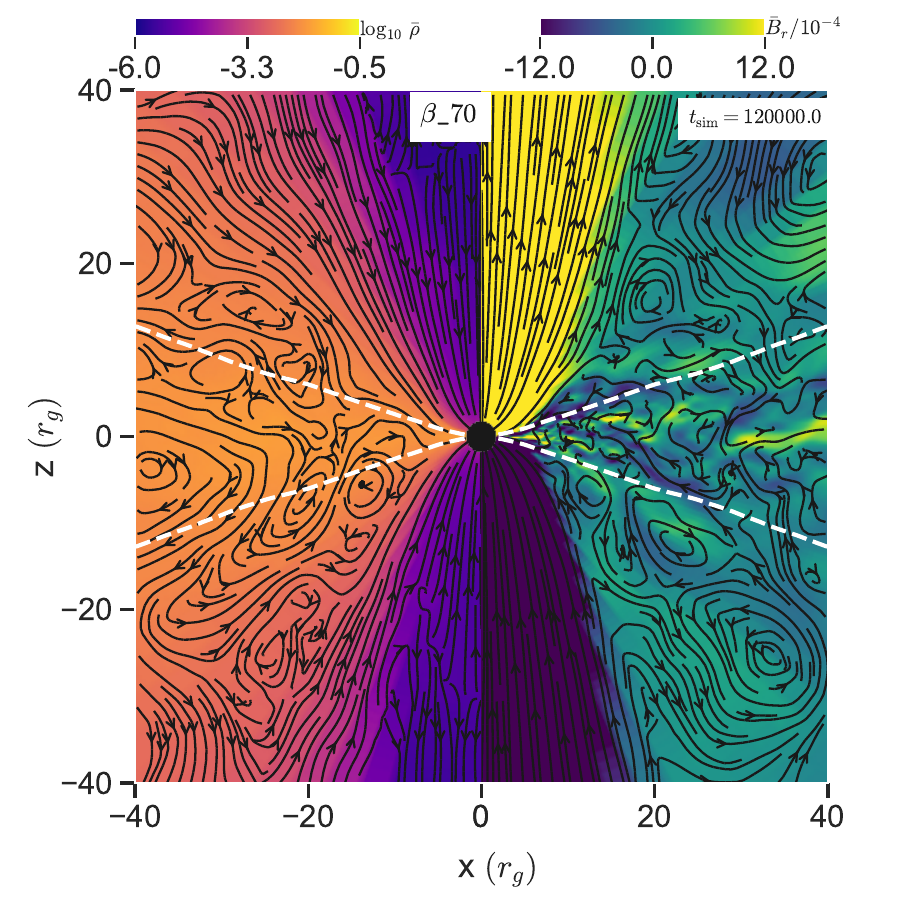}{0.49\textwidth}{(d)}
     }     
\caption{A zoomed-in view of the flow structures for the four representative runs: Initial RIAF, $\beta\_7000$, $\beta\_1500$ and $\beta\_70$ at the end of simulation run time. The left panels in each Fig. show the $\phi$-averaged rest mass density ($\bar{\rho}$) and poloidal velocity streamlines, while the right panels in each Fig. show mean radial field $\bar{B}_r$ (color) and poloidal magnetic field lines ($\bar{B}_{p}$) (streamlines).  }
\label{fig:flow_zoom_in}
\end{figure*}

The qualitative pictures discussed above are representative of all our simulations. A close-in view of flow and magnetic field structures at late times for this three representative runs along with the Initial RIAF run are shown in Fig. \ref{fig:flow_zoom_in}. The snapshots show that most of the advected flux is concentrated in the low-density laminar funnel region (polar region) with a different sign across the mid-plane. At the same time, the turbulent disk mid-plane has small patches of the magnetic field of both polarities. The animations of the other runs with external magnetic field loops, along with the Initial RIAF run, can be viewed in this \href{https://www.youtube.com/watch?v=3GyW2jRDbcU&list=PLUKo6vYd0sPJ69kl5JdWoBdzXAI84o7V4}{YouTube playlist}. In the upcoming sub-sections, we will quantify different metrics of magnetic flux transport in greater detail for all the runs we performed listed in Table \ref{tab:loop}.

\subsubsection{Evolution of magnetic flux}
\begin{figure*}
%\centering
\gridline{\fig{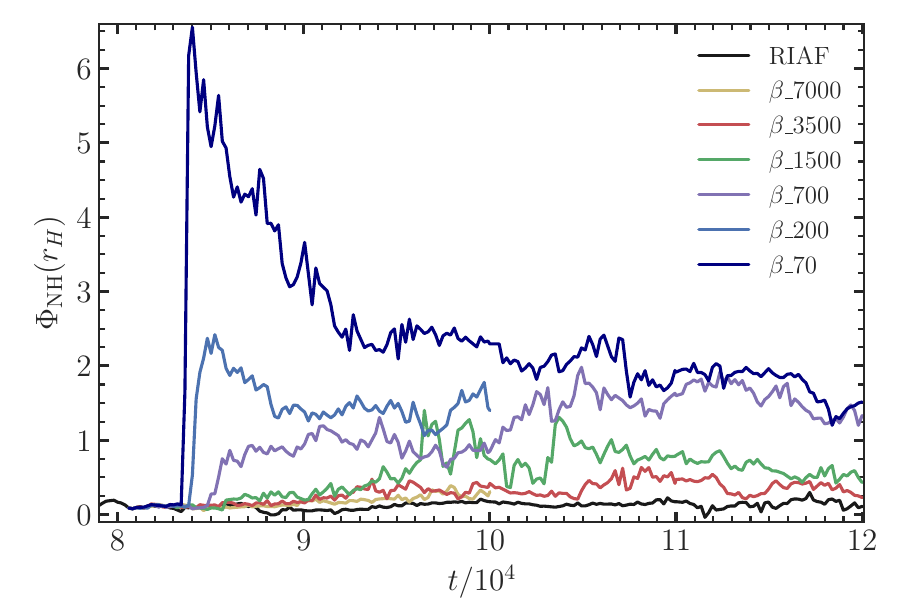}{0.49\textwidth}{(a)}
          \fig{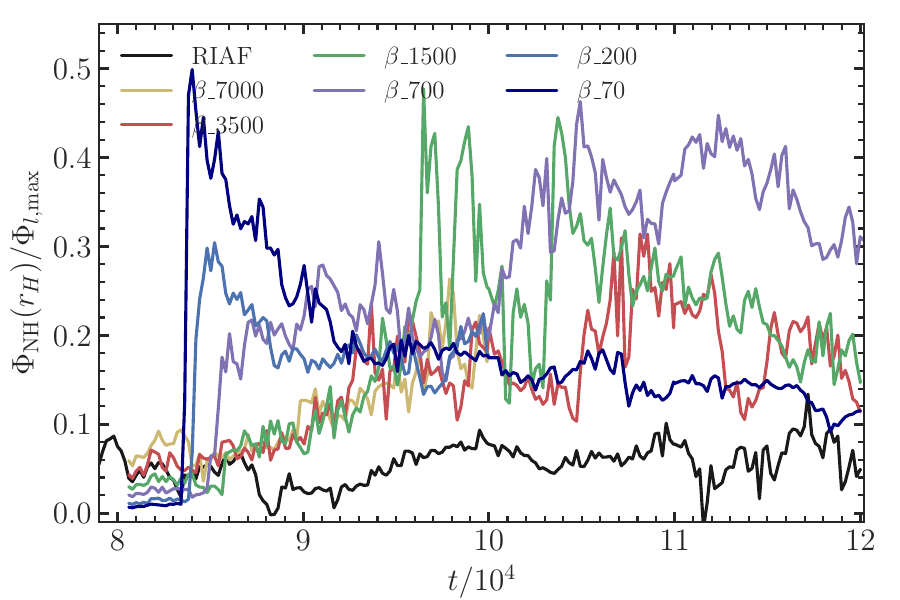}{0.49\textwidth}{(b)}
          }
\caption{Left panel: Comparison of the amount of radial magnetic flux threading the event horizon in the northern hemisphere $\Phi_{\rm NH} (r_H) $ for the runs with injected loops of different plasma $\beta$. Right panel: Fraction of the injected magnetic flux reaching the BH for runs with different plasma $\beta$.}
\label{fig:flux_r_strength}
\end{figure*}

Fig. \ref{fig:flux_r_strength}(a) shows the time evolution of the magnetic flux through the event horizon in the northern hemisphere, $\Phi_{NH} (r_H)$ for runs with different strengths of injected magnetic field loops, which are also compared with that of the Initial RIAF run. It is clearly visible that the injection of an external loop enhances the amount of flux at the event horizon.
In the runs with low $\beta_l\lesssim1000$, there is a transient rise of $\Phi_{\rm NH}$ due to the fast transport from the strong MRI channel flow. In the more extreme case of $\beta_l=70$, the transient phase is so extreme that it leads to a strong initial spike in $\Phi_{\rm NH}$, followed by a gradual decline towards a more steady flux level. For other runs with $\beta_l\gtrsim1000$, the build up of magnetic flux in the BH horizon is more gradual, and the build up is slower for runs with with higher $\beta_l$. It is worth noting that none of our simulations with injected loops reach the MAD state, with MAD parameter ranging from $\phi_{BH}=2$ to $\phi_{BH}=10$.

\begin{figure*}
\centering
 \includegraphics[scale=0.6]{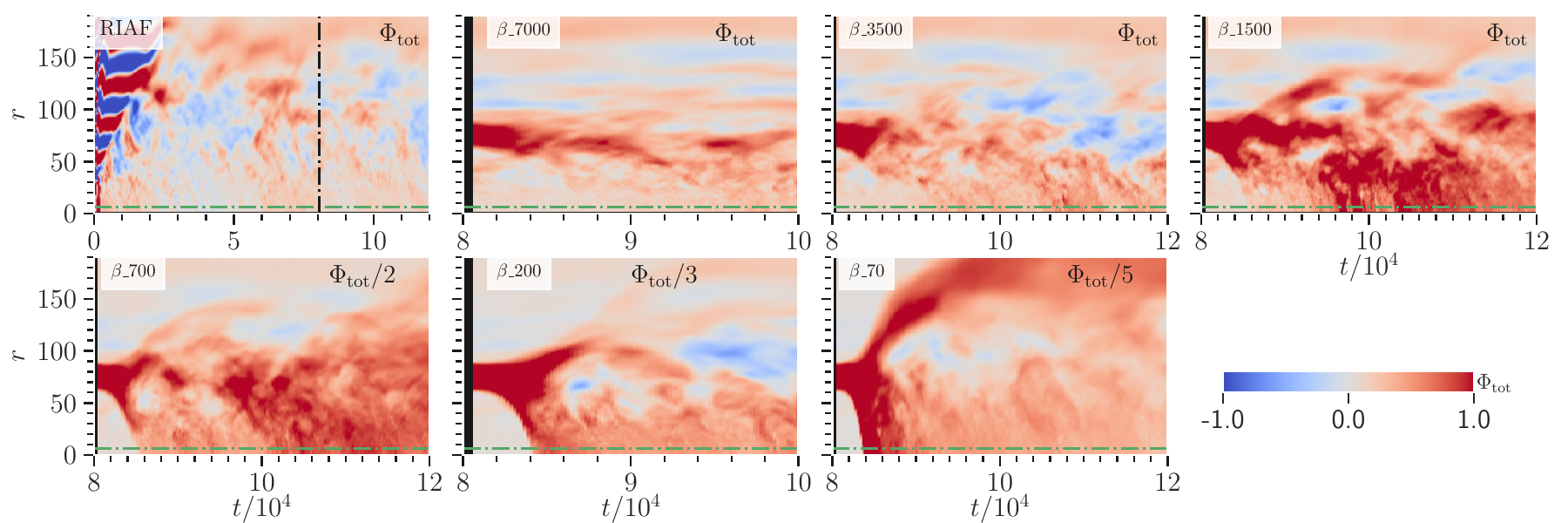}
\caption{Space-time plot for $\Phi_{\rm tot}$ defining the amount of poloidal magnetic flux at the disk midplane. The horizontal green dashed line shows the location of ISCO. The vertical black dashed line in the top left panel (the Initial RIAF) denotes  the time of injection of external flux.}
\label{fig:flux_mid_rt_strength}
\end{figure*}

We show the spatial-temporal variation of total flux $\Phi_{\rm tot} (r,t)$ available for accretion in the northern hemisphere in  Fig. \ref{fig:flux_mid_rt_strength} to obtain a more complete picture of the flux transport at different radii. Each panel of  Fig. \ref{fig:flux_mid_rt_strength} 
describes the evolution of the radial profile of $\Phi_{\rm tot}$ over time for runs with different $\beta_l$. The first panel in the top row corresponds to the Initial RIAF run.
It again demonstrates that the system forgets its initial magnetic field configuration after the time $3-4\times10^4$. The rest of the panels show the spatial-temporal evolution of $\Phi_{\rm tot}(r,t)$ for other runs after we inject external magnetic field loops.
In accordance with Figure \ref{fig:flux_r_strength}, we see that 
there are two regimes of flux transport depending on the strength of the injected loop. With very strong external flux ($\beta_l=70, \ 200$), the external flux is quickly transported both inwards and outwards, characteristic of the
channel flows with flow directions alternating over height, as seen in the last tow bottom panels of Fig. \ref{fig:flux_mid_rt_strength}. The channel flows lead to an initial transient transport of a large fraction of initial flux into the BH, followed by subsequent relaxation and diffusion towards a more steady flux level.

With weak external flux ($\beta_l \gtrsim3500$), the initial external magnetic flux gradually diffuses while being advected inwards.
In the end, a fraction of the flux overcomes diffusion to reach the BH, which will be discussed more quantitatively in the next subsection. 
In between  these two regimes, there lies the case of moderately strong external flux ($\beta_l=700, \ 1500$), for which the flux transport by the channel flows diffuses before reaching the BH, and subsequent transport is likely mediated by a combination of advection and diffusion.

 \subsubsection{The Efficiency of transport}
 
Till now, we considered the total flux as the primary diagnostics irrespective of the amount of flux associated with the injected loops. However, it is worth noting that different loops have different amounts of fluxes. Therefore, a normalised flux $f_B$ as defined in equation \ref{eq:efficiency} would be the better indicator of the efficiency of flux transport. 
 
Fig. \ref{fig:flux_r_strength}(b) shows the time variation of the fraction $f_B$ for different runs. In the regime of very strong injected flux ($\beta_l=70, \ 200$), the efficiency is quite high (up to $\sim50\%$) during the initial phase when channel flows dominate. Later, the efficiency goes down to around 15-20$\%$.
In the weak field regime ($\beta_l \geq 3500$), despite that flux in the BH is accumulated gradually, the efficiency of flux transport is more or less similar,
which is around 15-20 percent. 
For comparison, we also show the result for the Initial RIAF run, where we define $\Phi_{l, {\rm max}}$ by calculating flux at a radius $r=75$ at $t=0$.
Finally, it is interesting to note that flux transport appears to be more efficient, reaching about 20-40$\%$, when the field strength is in between the two regimes, i.e for $\beta_l=700$ and $\beta_l=1500$. 
  
\subsubsection{Effects on mass accretion rate and accretion stresses}
\label{sect:mdot_strength}
 
 \begin{figure}
      \centering
      \includegraphics[scale=0.55]{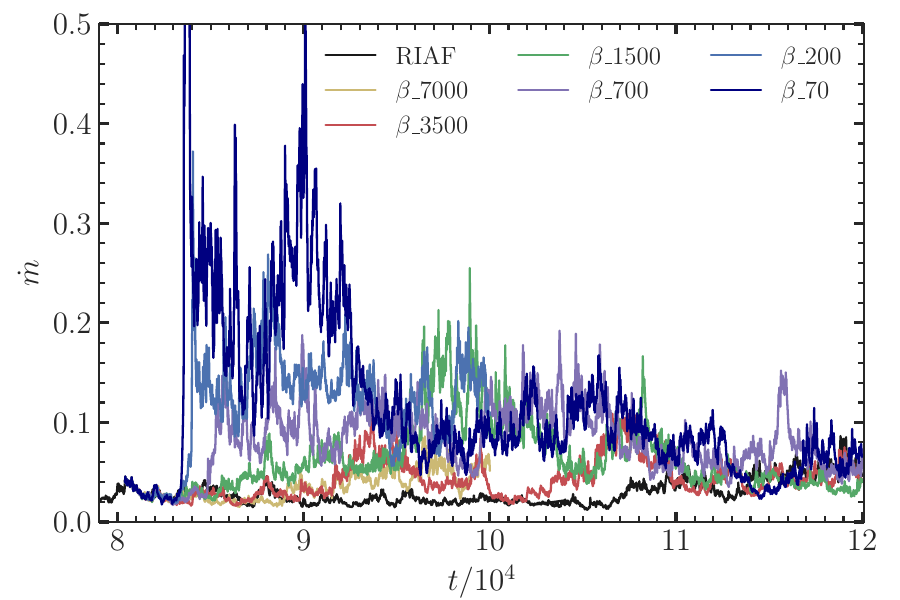}
      \caption{Time evolution in mass accretion rate $\dot{m}_{\rm in}$ at the event horizon after the injection of magnetic field loops of different strengths defined by plasma $\beta_l$.}
      \label{fig:mdot_strength}
 \end{figure}
 
 \begin{figure*}
\includegraphics[scale=0.6]{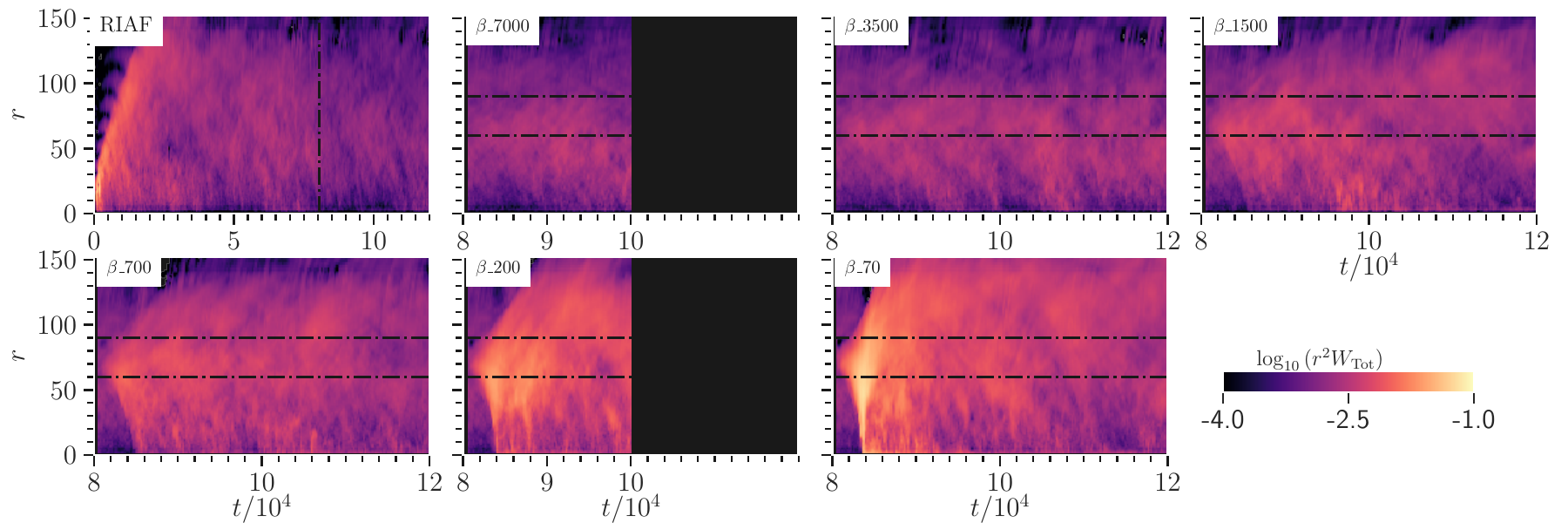}
%\gridline{\fig{Wmax_rt_strength_rout_152.pdf}{0.95\textwidth}{(a)}
%          }
%\gridline{\fig{Wrey_rt_strength_rout_152.pdf}{0.95\textwidth}{(b)}
%          }
\caption{Space-time plot of the radial profile of the average midplane total accretions tress $\la W_{\rm Tot} \ra = \la W_{\rm Max} \ra  + \la W_{\rm Rey} \ra$ for different runs, where the average is done in the azimuthal and vertical directions within one-scale height from the midplane.
%Here, $W_{\rm Max}$ and $W_{\rm Rey}$ are Maxwell and Reynolds stresses respectively. 
The vertical black dashed line in the top left panel indicates the injection time of field loops. Horizontal black dashed lines in the rest of the panels mark the radial range where field loops are injected (between $r=60$ and $r=90$).}
\label{fig:acc_stress}
\end{figure*}

In this subsection, we study how the injection of external magnetic flux influences the accretion properties such as the mass accretion rate and accretion stresses. Fig. \ref{fig:mdot_strength} shows time history of mass accretion rate at the event horizon for runs with a range of $\beta_l$. We further show in Fig. \ref{fig:acc_stress} the space-time plot of the azimuthally and vertically (over one scale-height) averaged total accretion stress $\la W_{\rm Tot} \ra$, which is a combination of 
Maxwell and Reynolds stresses defined in the orthonormal fluid frame (see section \ref{sect:convergence}) as,
\bea
&& W_{\rm Max}  = 2 p_{\rm mag} \ u^{\hat{r}} \ u^{\hat{\phi}} - b^{\hat{r}}\ b^{\hat{\phi}}, \\
&& W_{\rm Rey} = \left(\rho + \frac{\gamma}{\gamma-1} p_{\rm gas} \right) u^{\hat{r}} \ u^{\hat{\phi}}, \\
&& \la W_{\rm Tot} \ra= \la W_{\rm Max} \ra+ \la W_{\rm Rey} \ra \ ,
\eea 
where the Maxwell stress is the dominant component.

A fresh injection of external field loops reignites the linear MRI and leads to higher accretion stresses and hence an increase in mass accretion rate. We find that with high field strength in the loop (i.e $\beta_l\lesssim200$), there is substantially enhanced accretion stress, leading to a rapid, strong and transient increase of accretion rate. The stresses are reduced after the transient phase but are still much stronger than those in the Initial RIAF run within simulation time, reflecting the prolonged influence of the initial flux loop.
Simulations with $\beta_l\gtrsim3500$ show only a modest increase of accretion stress and the accretion rates compared to the Initial RIAF run, indicating that the injection of external flux has only a minor impact on disk turbulence.  
For simulations with intermediate $\beta_l$, there is a modest enhancement of the accretion stress, resulting in a modest enhancement of the accretion rate. We also note that after $t=1.1 \times 10^5$, despite having a higher level of magnetic flux (compared to the Initial RIAF run; Fig. \ref{fig:flux_mid_rt_strength}), mass accretion rate in the runs with injected loops are very similar to that in the Initial RIAF run. This happens due to the quick depletion of mass supply in the disk at earlier times due to the enhanced stresses in the runs with injected loops.

\subsubsection{Disk and flow structures}
\begin{figure*}
 \gridline{\fig{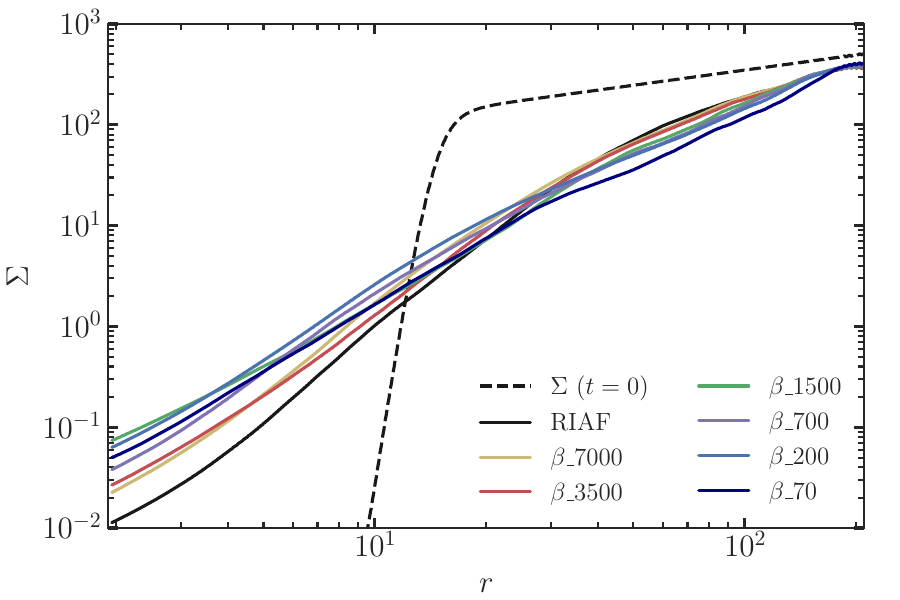}{0.49\textwidth}{(a)}
            \fig{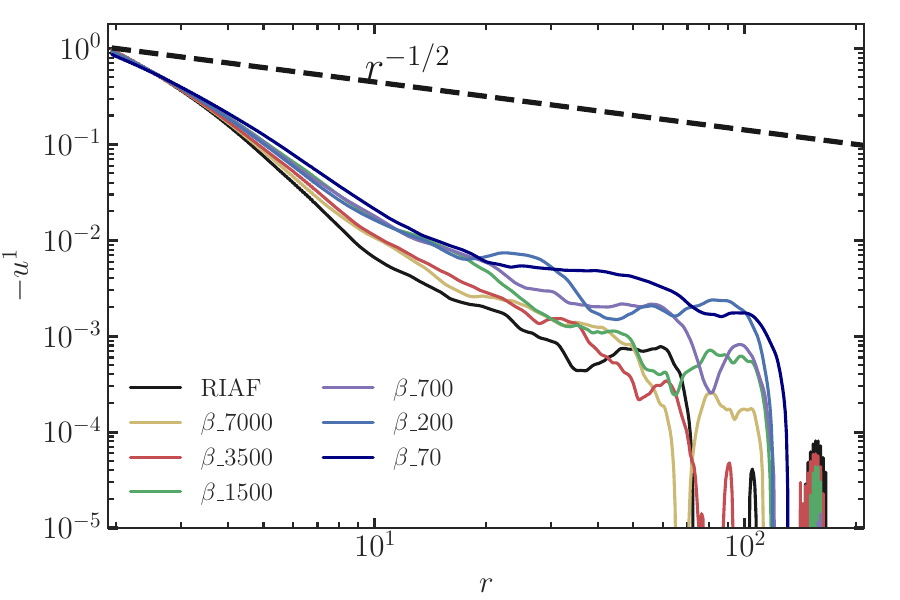}{0.49\textwidth}{(b)}
 }
\caption{Radial profiles of surface density ($\la \Sigma \ra$, left) and radial velocity ($\la u^1 \ra$) for the Initial RIAF run and runs with injected loops with fiducial size and location with different strengths.
%A slight difference between the surface density in the RIAF and that in other runs with injected loops implies that the flow dynamics did not change due to the injection of external loops. 
The time average is done over $t=9\times 10^4-10^5$.
%Right panel: Radial profiles of velocity $u^1$ 
For $\la u^1 \ra$, the spatial average is done over the azimuth and within one scale-height about the midplane.}
\label{fig:surf_den}
\end{figure*}

In this subsection, we further examine how external magnetic flux changes the disk structure and flow properties. We start by considering the radial profiles of surface density, defined as\footnote{We note that the standard definition (\ref{eq:Sigma}) asymptotes to $\rho r^2d\theta$ at large radii, with an extra $r$ factor compared to the Newtonian definition (assuming surface density is defined by integrating along spherical shells).}
\be
\Sigma(t,r) = \frac{1}{\phi_{\rm ext}} \int_{\phi=0}^{\phi_{\rm ext}} \int_{\theta=0}^{\pi} \rho \ \sqrt{-g} \  d \theta \  d\phi\ ,\label{eq:Sigma}
\ee
and the radial velocity $\la u^1 (r) \ra$, averaged within one scale height about the midplane. The results for the Initial RIAF run and runs with external field loops, time average is done over $t=9\times 10^4-10^5$, are shown in Fig. \ref{fig:surf_den}.

In the Initial RIAF run, we see that the accretion velocity approaches the free-fall velocity ($v_{ff} = \sqrt{2/r}$) within the ISCO, while accretion velocity ranges between $0.01-0.5$ of the Keplerian velocity further out till the radius of inflow equilibrium. Upon imposing an external field, the higher accretion stresses lead to higher accretion velocities. The enhancement can be up to a factor of $\sim10$ in the strong field case with $\beta_l=70$ at the representative radii of $r\sim30-60r_g$, while for weak field runs (e.g., $\beta_l\gtrsim3500$), the accretion velocity is only enhanced by a modest factor of $\sim2$. We also note that the profile of $\la u^1(r) \ra$ also evolves over time accompanying magnetic flux transport, but qualitatively, the profiles shown in Figure \ref{fig:surf_den} are representative over the duration of our simulations.

The altered accretion velocity profile $\la u^1(r) \ra$ further modifies the surface density profile. Generally, after imposing an external field loop, the surface density becomes steeper compared to the surface density profile in the Initial RIAF run, though the deviation is only modest. The surface density profile also evolves over time.
We note that earlier RIAF simulations of the SANE state already indicated that there might not be any universal power law for the surface density profile and other flow properties \citep{White2020}. When supplied with external magnetic flux in the outer disk, our results suggest additional surface density variations during the process of magnetic flux transport. In other words, the dynamics of RIAFs are dependent on the magnetised mass reservoir at larger radii.

\subsection{Results for the other parameters}

\begin{figure*}
\gridline{\fig{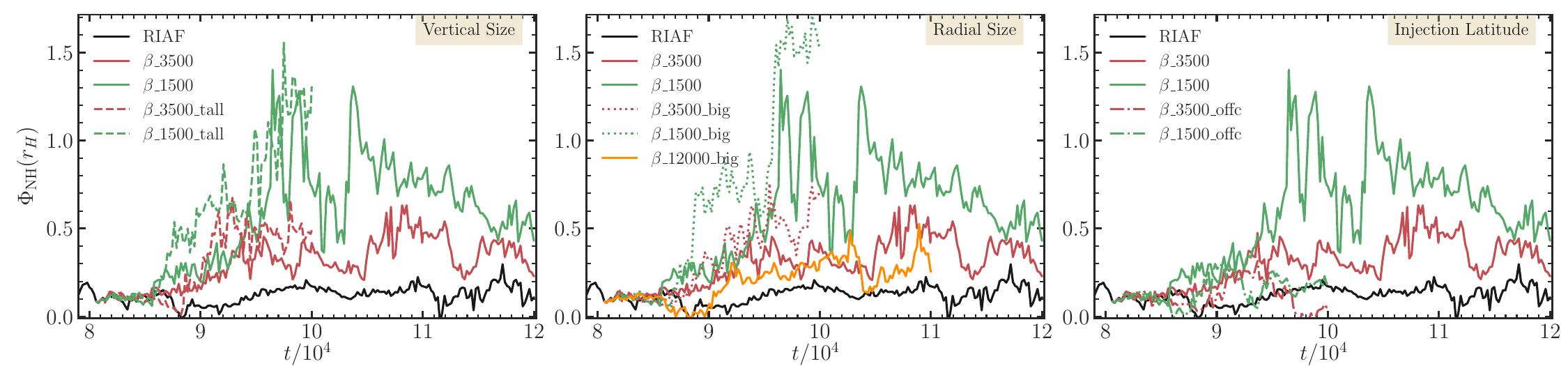}{0.95\textwidth}{(a)}
          }
\gridline{\fig{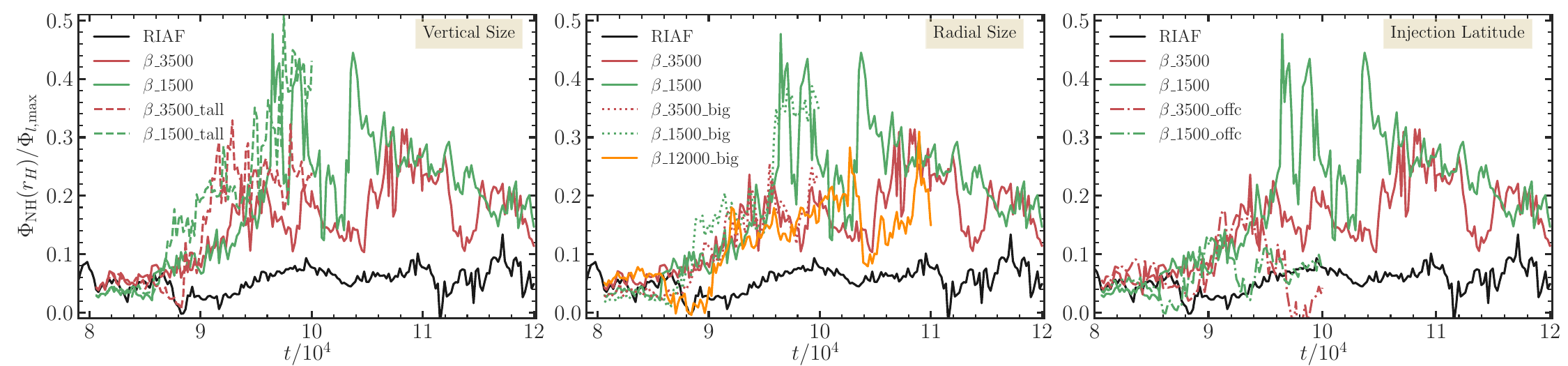}{0.95\textwidth}{(b)}
          }
\caption{The time evolution of the amount of radial magnetic flux threading the event horizon in the northern hemisphere $\Phi_{\rm NH}(r_H)$ (top panels) and its fraction over the injected magnetic flux (bottom panels) for the runs with injected loops of larger vertical size (left panel), larger radial size (middle panel), and off-centered injection latitudes (right).
The results are compared with runs with fiducial loop geometries and the Initial RIAF run with the same level of magnetization ($\beta_l=1500$ and $3500$).}
\label{fig:flux_r_size}
\end{figure*}

In this section, we assess the robustness of our fiducial simulation results by considering different geometries for the injected field loops. In particular, we change loop sizes (both vertical and radial) and the injection latitudes. We focus on loops of fiducial strength $\beta_l=3500$ and $\beta_l=1500$, respectively. Additionally, we study the transport of a big loop of similar strength ($\beta_l=12200$) to mean poloidal fields produced by MRI dynamo in the Initial RIAF run in the quasi-stationary phase. Fig. \ref{fig:flux_r_size} shows the time evolution of the radial magnetic flux threading the event horizon in the northern hemisphere $\Phi_{\rm NH} (r_H)$ (top panels) and flux transport efficiency (bottom panels) for these additional simulations. 
  
\subsubsection{Vertical and Radial sizes}

The left panels of Fig. \ref{fig:flux_r_size} compare the flux transport for taller loops of vertical size $z_l =2.5H$ with the loops of similar strength but of fiducial size $z_l=1.5H$. We find that the vertical size of the loops does not affect the amount of flux reaching the BH, and the efficiency of flux transport remains almost unaltered with the change of the vertical size of the loops.
 
Similarly, the middle panels of Fig. \ref{fig:flux_r_size} compare flux transport
between loops of different radial sizes, where we consider bigger loops of radial size $\Delta r_l=60$ as opposed to the fiducial radial size of $\Delta r_l=30$. We observed that a larger amount of flux reaches the BH for the bigger loops, which is reasonable because more magnetic flux is available in these loops compared to its smaller counterparts.
However, the fraction of the flux reaching the BH remains similar for both the smaller and bigger loop cases with the same plasma $\beta_l$. This result also holds for our additional run with $\beta_l=12200$. This indicates that the efficiency of flux transport remains unaffected by the radial extent of the injected loops.

\subsubsection{ Injection latitude}

In addition  to studying the effects of strength and size of the loops on the transport process, we also consider injecting off-centred loops with plasma $\beta_l=3500$ and $\beta_l=1500$ to examine whether loop injection away from the mid-plane facilitates flux transport or not. The comparison with our fiducial injection prescription is shown in the right panel of Fig. \ref{fig:flux_r_size}.
Surprisingly, the injection of off-centred loops leads to a distinctly lower flux level at the event horizon. 
While it is not entirely clear why this is the case, we speculate that it is related to stronger magnetic reconnection in the off-centred case that leads to more considerable destruction of magnetic flux, which occurs during the interplay between the injected field and the dynamo-generated background field.
 
The stark difference between the magnetic field evolution in off-centred and the fiducial cases can be also seen by comparing the movies describing the magnetic field loop evolution for the runs beta\_1500 (\href{https://www.youtube.com/shorts/duxJRtFO0zI}{movie-beta-1500}) and beta\_1500\_offc (\href{https://www.youtube.com/shorts/y0lweEVgfGI}{movie-beta-1500-offc}) respectively.

\section{Discussion}

\subsection{Inefficiency of Dynamo in SANE/RIAF}
We threaded the initial geometrically semi-thick disk ($H/R\approx 0.2$) with small magnetic 
field loops of alternating polarity and attained a quasi-stationary weakly magnetized RIAF (SANE; see Fig. \ref{fig:jnet_mad_param}), that does not remember the initial field geometry (see sections \ref{sect:flow_evo_riaf} and \ref{sect:sane_or_mad}). An MRI dynamo is responsible for generating and sustaining magnetic fields (both small-scale and large-scale) in the quasi-stationary RIAF (\citealt{Hawley2013,Hogg2018}). A large-scale dynamo does operate (\citealt{Dhang2019}) and generate large-scale magnetic fields in the weakly magnetized RIAF (see last two panels of Fig. \ref{fig:flow_evo}), but not efficient enough to produce strong magnetic fields that can convert a SANE to MAD. This result aligns with earlier works which found dynamo action in a SANE RIAF does not lead to jet formation (\citealt{Beckwith2008,Narayan2012}). Earlier works (\citealt{Hogg2018, Dhang2020}) with different numerical set-ups  investigating MRI dynamo lead to the conclusion that dynamo action is weak in a geometrically thick RIAF. Hence it is likely that the inefficiency of jet formation is likely to be attributed to insufficient poloidal field generation (weak $\alpha$-effect) and strong turbulent pumping which transports large-scale magnetic field  radially outward in a RIAF(\citealt{Dhang2020}). 

Recently, \citet{Liska2020} reported that when starting the simulation with an unusually strong ($\beta \approx 5$) and coherent toroidal magnetic field, the MAD state can be achieved at late times. They argued that an MRI dynamo could produce strong poloidal field loops of size $H \propto R$ from the very strong and coherent initial toroidal field. The further the creation location is, the bigger the loops are. Most of the loops move outward, while a few `lucky' loops created at large radii are somehow arrested and stretched inward and lead to the MAD state.
However, how the accretion disk can possess such a coherent initial toroidal field of the same polarity spanning several decades in radii at the first place remains questionable.

Overall, we reaffirm that the MRI dynamo in the standard SANE state does not spontaneously generate a strong coherent large-scale poloidal field to turn the disk into the MAD state. In the absence of an initial poloidal field, achieving the MAD state may require an unusually strong and coherent toroidal field that may be unpractical in reality.

\subsection{Plausible sources of external magnetic fields}
\label{sect:source_loop_discuss}
In this work, we considered the possibility that the disk acquires an external poloidal field in the form of field loops of different sizes and shapes. 
What can be the source of such external field loops? While definitive evidence is lacking, we speculate that accreting such external field loops is plausible in a variety of systems.

In the hard state of the XRBs, a RIAF close to the BH is proposed to be connected to an outer thin disk \citealt{Esin1997, Done2007}, which can supply large-scale magnetic flux to the inner RIAF. The outer thin disk can in principle harbour large-scale 
magnetic field due to an efficient dynamo action (\citealt{Flock2012a, Gressel2015}) or due to the coronal accretion of magnetic flux (\citealt{Guilet2012}) from the companion/donor star, or a combination of both.
The donor star in the low mass XRBs are likely to be either K or M-type dwarf stars (\citealt{Fragos2015}) or evolved stars (e.g., as in GRS 1915+1105). The donor stars in the XRBs are supposed to be tidally locked to the rotation period of the binaries with an orbital period of hours to days (\citealt{Coriat2012}). These fast-rotating dwarf stars show vigorous magnetism with a surface magnetic field of strength ($\sim 10^3$ G) similar to sunspots (\citealt{West2008,Davenport2016}). Additionally, in the active region, the magnetic field is one order of magnitude stronger than the average stellar magnetic field. Magnetized matter from the donor star passes through the first Lagrange point ($L_1$) and enters the Roche lobe of primary (accretor) almost ballistically and circularizes at the circularization radius (\citealt{Frank2002}).
We speculate that the mass loss from the L1-nozzle may proceed through a chain of mass blobs encircled by field loops (e.g., as also considered in \citealt{Ju2017}), which may get amplified and become quasi-axisymmetric during the circularization process.
Thus, if this external flux can be brought in through the outer thin Keplerian disk, then it may further feed the inner RIAF, where flux transport is efficient and saturate the BH.

The accretion flow in a low-luminosity AGN is also thought be a RIAF. In this case, the gas supplied by the ambient medium to the accretion flow is magnetized. It can harbour a large-scale magnetic field as inferred from the observation of the large-scale poloidal flux in the Galactic centre (\citealt{Nishiyama2010}). Recent numerical simulations by \citet{Ressler2020a,Ressler2020b} found that the large-scale accretion flow around the galactic centre fed by the winds of Wolf-Rayet stars can achieve the MAD state, with efficient inward transport of magnetic field embedded in the accreting material. Their injected magnetic fields have a pure toroidal component with random orientation, thus we may consider that such fields effectively enter the accretion disk in the form of closed field loops from some random directions. Our results are in line with their findings, while our controlled experiments further provide a physical basis for better understanding the efficient flux transport around SMBHs.

\subsection{Transport efficiency in the SANE and its possibility of transformation to MAD}
 In section \ref{sect:results_flux}, we show the results of the effects of the loop injection in turbulent quasi-stationary SANE RIAF. We observed a simultaneous transport of magnetic flux inward  and outward due to the channel flows (more evident in the strong field case in Fig. 8 and 10). For strong field case, channels quickly transport the field inward giving very high efficiency, for weak field cases, a fraction of flux reaches BH slowly overcoming diffusion. We found that except for off-centered loops, transport of externally injected magnetic flux loops is relatively efficient, with typically $\sim20\%$ of the available flux end up being accreted to the central BH regardless of initial field strength and size.

Note that, we find all of the simulations with injected loops  have the MAD parameter $\phi_{BH} \leq 10$.  This implies that none of our simulations reaches the MAD state, but if this relatively high efficiency of flux transport obtained from our controlled experiments is universal, we can estimate the requirement on the external flux to potentially transform a SANE disk into MAD.

The MAD parameter $\phi_{\rm BH}$ (the normalised unsigned flux threading the BH) is related to the magnetic flux (signed) $\Phi_{NH}$ threading the northern hemisphere of the BH as follows 
\be
\phi_{\rm BH} \approx \frac{\Phi_{NH} (r_H)}{\sqrt{\dot{m}}r_g c^{1/2}}.
\ee 

We have found that
a certain fraction $f_B = \Phi_{NH} (r_H)/\Phi_{in}$ (equation \ref{eq:efficiency})
of the injected flux $\Phi_{\rm in}$ reaches the BH. Earlier numerical experiments suggest that the MAD could be achieved if the MAD parameter exceeds a critical value $\phi_{BH,c}$ at the event horizon (e.g., see \citealt{Tchekhovskoy2011}). This indicates to 
a critical value of the injected flux $\Phi_{in,c}$ which is a plausible minimum flux required for the MAD state and it is given by
\be
\Phi_{\rm in,c} = \left(\frac{r_g c^{1/2}}{f_B} \right) \phi_{{\rm BH},c} \sqrt{\dot{m}}.
\ee 
Therefore,  the minimum poloidal magnetic field required at the injection location is given by
\be 
B_{\rm in,c} = \left(\frac{r_g c^{1/2}}{2 \pi f_B} \right) \left(\frac{r}{\Delta r} \right) \left( \frac{\phi_{{\rm BH},c}} {r^2} \right) \sqrt{\dot{m}},
\ee 
where we have estimated that for a loop centered on radius $r$ with half-width $\Delta r$, $\Phi_{\rm in,c}\approx 2\pi B_{\rm in,c} r\Delta r$. If we take $\Delta r/r=0.2$, then the value of $B_{\rm in,c}$,
in terms of Eddington accretion rate $\dot{M}_{\rm Edd}=1.5 \times 10^{19} \ M_{10} \ {\rm gm} \ {\rm s}^{-1}$ is given by
\be
B_{\rm in,c} \approx  \frac{10^{4}}{f_B} \ \phi_{40} \   r_{100}^{-2} \ \dot{m}_{-4}^{1/2} \ M_{10} ^{-1/2} \ G,
\ee
where $\phi_{40}=\phi_{{\rm BH},c}/40$, $r_{100}=r/(100r_g)$, $\dot{m}_{-4} = \dot{m}/10^{-4}\dot{M}_{\rm Edd}$, and $M_{10}=M_{BH}/10M_{\odot}$.

Would this amount of magnetic field be available for accretion at the outer radii of the RIAF? We will estimate the poloidal field strength available for accretion in case of an XRB. In the low hard state, the RIAF close to the BH is proposed to be connected to an outer thin disk. The plausible source of the large-scale magnetic field in the thin disk could be the dynamo action. Another scenario would be the advection of large-scale field loops from the companion star as discussed in section \ref{sect:source_loop_discuss}. Independent of the mechanism, we can estimate the characteristic poloidal field strength in the thin accretion disk given that the accretion is driven by radial transport of angular momentum in the disk as (e.g., \citealt{Bai2009})
\be
-\overline{B_r B_{\phi}} \approx \frac{\dot{m} \Omega}{h_{a}}.
\ee
Here, $h_a=\xi H_{\rm thin}$ is the thickness of the disk over which accretion proceeds. Further, if we assume that $|B_r|\approx 1/5 |B_{\phi}|$ (which is found to be consistent in the MRI simulations of accretion disks) and $\xi = 6$, then the total radial magnetic field in the thin disk of aspect ratio $\epsilon_{\rm thin} = H_{\rm thin}/R$ is given by, 
\be
\label{eq:Br_thin_disk}
B_{r,d} \approx 10^4 \ \epsilon^{-1/2}_{0.05}   \dot{m}_{-4}^{1/2} \ r_{100}^{-5/4} \ M_{10}^{-1/2} \ G,
\ee
where $\epsilon_{0.05}=\epsilon_{\rm thin}/0.05$. It is to be noted that in the strongly magnetized coronal region of the thin disk, a large share of this estimated total field $B_{r,d}$ will likely be in the mean coherent part of the magnetic field.
In reality, the mass accretion rate in the outer thin disk is expected to be higher compared to that in a RIAF (\citealt{Yuan2014}), and hence the total poloidal field will also be higher. Therefore, comparison of $B_{{\rm in,c}}$ and $B_{r,d}$ leads to the inference that it is quite possible that in an XRB, the outer thin disk reservoir can potentially supply an adequate amount of magnetic flux to the inner RIAF that eventually may form a MAD close to the BH.

\section{Summary}
In this paper, we studied the magnetic field generation and transport in a geometrically thick RIAF. We initialize the disk with  magnetic field loops of alternate polarity so that the quasi-stationary RIAF is weakly magnetized, i.e in the SANE regime. In this quasi-stationary turbulent SANE RIAF, we study the transport of external magnetic flux (in the form of loops) of different strengths, sizes and shapes. Here we outline the key findings of our work.
\begin{itemize}
    \item We reconfirm that the MRI dynamo in a standard SANE RIAF does not generate a strong coherent large-scale poloidal field to turn the SANE state into the MAD state. 
    \item Magnetic flux transport is relatively efficient in the SANE RIAF: fifteen to forty percent of the external magnetic flux injected at the outer radii is able to reach the BH.
    \item Flux transport efficiency is independent of the loop parameters such as strength and size. However, if the loops are injected at high latitudes rather than at the mid-plane, the efficiency becomes poor.
    
\end{itemize}

We also find that accretion flow profiles (e.g surface density, accretion velocity) are altered as external magnetic flux is injected into the disk. We propose that the dynamics of the RIAF depend on the magnetized mass reservoir at the outer radii.

Based on our results, we argue that it might be easier to transform a SANE disk to a MAD by supplying external poloidal field loops at the outer disk provided that the relatively high efficiency of flux transport obtained from our controlled experiments is universal.

It is to be noted that as a first study, it is not yet clear which factors determine the magnetic flux transport efficiency of $\sim 15-40 \%$ in our work. Additionally, we must mention that we have studied the transport of external magnetic flux in the quasi-stationary turbulent RIAF in limited parameter space. For example, we have considered only one injection location with  the inner edge of the loop being at $r=60$,  whereas, in reality, the loops are supposed to be available for accretion as far as in the disk truncation region in XRBs, or even at larger radii in Low-luminosity AGNs.
We plan to explore magnetic flux transport with different configurations and with larger dynamical ranges.
Furthermore, future work should extend this study to the thin disk regime, which is applicable to regions beyond the truncation radius in the low/hard state of the XRBs, as well as in luminous AGNs.

We thank Ramesh Narayan for his initial input into this project. We also thank Kandaswamy Subramanian and the anonymous referee for constructive suggestions. This research was supported by NSFC grant 11873033. Numerical simulations are conducted on TianHe-1 (A) at the National Supercomputer Center in Tianjin, China, and on the Orion cluster at the Department of Astronomy, Tsinghua University.

All the movies of the simulations mentioned in Table \ref{tab:loop} are available in this \href{https://www.youtube.com/watch?v=3GyW2jRDbcU&list=PLUKo6vYd0sPJ69kl5JdWoBdzXAI84o7V4}{YouTube link}.

\bibliography{bibtex_tran}{}
\bibliographystyle{aasjournal}



\end{document}